\documentclass[12pt,epsf,qsymbols]{article}
\pdfoutput=1
\usepackage[utf8]{inputenc}
\usepackage[normalem]{ulem}
\usepackage[english]{babel}
\usepackage{tabularx}
\usepackage{array}
\usepackage{graphics}
\usepackage{graphicx}
\usepackage{psfrag}
\usepackage{epsfig}
\usepackage{amsmath}
\usepackage{amssymb}
\usepackage{bbold}
\usepackage{slashed}
\usepackage{ulem}
\usepackage{setspace}
\usepackage{rotating}
\usepackage{colortbl}
\usepackage{longtable}
\usepackage{slashed}
\usepackage{braket}
\usepackage{lineno}
\makeatletter
\usepackage{textcomp}
\usepackage[usenames,dvipsnames]{xcolor}
\usepackage{relsize}
\usepackage{cite}

\usepackage{bbm}
\usepackage{bm}
\usepackage{enumerate}
\usepackage{amsmath}
\usepackage{verbatim}

\usepackage{feynmp-auto}

\usepackage{hyperref}
\hypersetup{colorlinks,citecolor=nicegreen,linkcolor=niceblue}
\hypersetup{colorlinks=true}

\setlongtables

\newcommand{\cb}{{\mathcal B}}
\newcommand{\bea}{\begin{eqnarray}}
\newcommand{\eea}{\end{eqnarray}}
\newcommand{\beq}{\begin{equation}}
\newcommand{\eeq}{\end{equation}}
\newcommand{\ec}{\end{center}}
\newcommand{\bc}{\begin{center}}

\newcommand{\gev}{{\rm GeV}}
\newcommand{\mev}{{\rm MeV}}

\newcommand{\pdir}{p\kern -5.2pt\raise 0.2ex\hbox {/}}

\newcommand{\vdir}{v\kern -5.75pt\raise 0.15ex\hbox {/}}
\newcommand{\kdir}{k\kern -5.75pt\raise 0.15ex\hbox {/}}
\newcommand{\epsdir}{\epsilon\kern -5.0pt\raise 0.15ex\hbox {/}}
\newcommand{\bvdir}{\bar{v}\kern -5.75pt\raise 0.15ex\hbox {/}}
\newcommand{\Ddir}{D\kern -7.75pt\raise 0.20ex\hbox {/}}
\newcommand{\Adir}{A\kern -7.75pt\raise 0.20ex\hbox {/}}
\newcommand{\ldir}{l\kern -5.0pt\raise 0.2ex\hbox{/}}
\newcommand{\varepsdir}{\varepsilon\kern -5.5pt\raise 0.15ex\hbox{/}}

\def \eff{{\text{eff}}}

\newcommand{\nn}{\nonumber}

\makeatother

\definecolor{niceblue}{rgb}{0.15,0.15,0.6}
\definecolor{nicegreen}{rgb}{0.1,0.5,0.1}
\definecolor{Red}{rgb}{1.,0.,0.}

\definecolor{Green}{rgb}{0.2,.7,0.2}

\usepackage{soul}

\begin{document}
\unitlength = 1mm

\thispagestyle{empty} 
\begin{flushright}
\begin{tabular}{l}
{\tt \footnotesize LPT-16-84}\\
{\tt \footnotesize IPPP/16/119}\\
\end{tabular}
\end{flushright}
\begin{center}
\vskip 3.4cm\par
{\par\centering \textbf{\LARGE  
\Large \bf Sterile neutrinos facing kaon physics experiments 
}}\\
\vskip 1.2cm\par
{\scalebox{.81}{\par\centering \large  
\sc A.~Abada$^a$, D.~Be\v{c}irevi\'c$^a$, O.~Sumensari$^{a,b}$, C.~Weiland$^{c}$ and R.~Zukanovich Funchal$^{b}$}
{\par\centering \vskip 0.7 cm\par}
{\sl 
$^a$~Laboratoire de Physique Th\'eorique (B\^at.~210)\\
CNRS and Univ. Paris-Sud, Universit\'e Paris-Saclay, 91405 Orsay cedex, France.}\\
{\par\centering \vskip 0.25 cm\par}
{\sl 
$^b$~Instituto de F\'isica, Universidade de S\~ao Paulo, \\
 C.P. 66.318, 05315-970 S\~ao Paulo, Brazil.}\\
{\par\centering \vskip 0.25 cm\par}
{\sl 
$^c$~Institute for Particle Physics Phenomenology, Department of Physics,\\ 
Durham University, South Road, Durham DH1 3LE, United Kingdom.}\\
{\par\centering \vskip 0.25 cm\par}

{\vskip 1.65cm\par}}
\end{center}

\vskip 0.85cm
\begin{abstract}
We discuss weak kaon decays in a scenario in which the Standard
Model is extended by massive sterile fermions.  After revisiting the
analytical expressions for leptonic and semileptonic decays we derive
the expressions for decay rates with two neutrinos in the final state.
By using a simple effective model with only one sterile neutrino,
compatible with all current experimental bounds and general
theoretical constraints, we conduct a thorough numerical analysis
which reveals that the impact of the presence of massive sterile
neutrinos on kaon weak decays is very small, less than $1\%$ on
decay rates. The only exception is $\cb (K_L\to \nu\nu)$, which can go
up to $\mathcal{O}( 10^{-10})$, thus possibly within the reach of 
the KOTO, NA62 and SHIP experiments. Plans have also been proposed to search for this decay at the NA64 experiment.
In other words, if all the future measurements of weak kaon decays
turn out to be compatible with the Standard Model predictions, this
will not rule out the existence of massive light sterile neutrinos with non-negligible active-sterile mixing.
Instead, for a sterile neutrino of mass below $m_K$, one might
obtain a huge enhancement of $\cb (K_L\to \nu\nu)$, otherwise
negligibly small in the Standard Model.
\end{abstract}
\newpage
\setcounter{page}{1}
\setcounter{footnote}{0}
\setcounter{equation}{0}
\noindent

\renewcommand{\thefootnote}{\arabic{footnote}}

\setcounter{footnote}{0}

\tableofcontents

\newpage

\section{Introduction}
\label{sec:intro}
The Standard Model (SM) predicts the strict conservation of lepton flavor
to all orders. The fact that neutrinos oscillate provides
clear evidence of the existence of physics beyond the Standard Model.
New physics models accommodating massive neutrinos and their mixing
open the door to many phenomena which basically have no Standard Model
background, such as lepton number violation, violation of lepton
flavor universality, or lepton flavor violation.

The experimental effort associated with the search of new physics
using observables involving leptons is impressive, and is currently
being pursued on all experimental fronts: (i) neutrino dedicated
experiments which aim to determine neutrino properties, such as the
Majorana/Dirac nature, the absolute neutrino masses, the hierarchy of
their mass spectrum, leptonic mixing and the \textit{CP}-violating phases; (ii)
high-intensity facilities that are studying several low-energy processes
 such as $\ell\to \ell'\gamma$, $\ell_k\to
\ell_i\ell_i\ell_j$, $\mu-e$ conversion in atoms, $\tau$ and meson $M$
decays ($\tau\to M\nu$, $M\to\ell\nu$, $M\to M'\ell\nu$, $M\to
M'\ell\ell, M'\nu\nu$, \dots); (iii) the Large Hadron Collider (LHC), which is the privileged
discovery ground of new particles, may also allow one to probe leptonic
mixing in the production and/or decay of the new states. On the other
hand, the recent cosmological data have put constraints on the sum of the light neutrino masses, 
which is especially restrictive when considering new light neutral states.

Among the several minimal possible scenarios, extending the SM with
sterile fermions -- which are singlets under the Standard Model gauge
group -- is a very appealing hypothesis, as their unique (indirect)
interaction with the Standard Model fields occurs through their mixing
with the active neutrinos $\nu_L$ (via their Yukawa couplings). Due to their
very unique nature, there is no bound on their
number, and \textit{a priori} no limits regarding their mass
regimes. Interestingly, sterile fermions (like right-handed neutrinos)
are present in many frameworks accounting for neutrino masses and the
observed mixing (as is the case for fermion seesaw mechanisms). The
interest in sterile neutrinos and their impact on observables strongly
depend on their masses.  Sterile neutrinos at the eV scales were proposed 
to solve neutrino oscillation
anomalies in reactors~\cite{reactor},
accelerators~\cite{Aguilar-Arevalo:2013pmq}, as well as in
  the calibration of gallium-target solar neutrino experiments~\cite{Giunti:2010zu}, all suggesting beyond
the three-neutrino paradigm.  The keV scale for sterile neutrinos offers
warm dark matter candidates~\cite{Asaka:2005an,Abada:2014zra}, and
explanations of some astrophysical issues such as kicks in pulsar
velocities~\cite{Adhikari:2016bei}.  Sterile fermion states with
masses above $10^9$ GeV have moderate motivation other than their
theoretical appeal (grand unified theories) and the possibility to have a
scenario for baryogenesis via the high-scale
leptogenesis~\cite{Davidson:2008bu}. Finally, the appeal for sterile
neutrinos in the range $\mev - \gev$ (and even TeV) strongly
resides in their experimental testability due to the many direct and
indirect effects in both high-energy (e.g. LHC)~\cite{Abada:2015zea}
and high-intensity (e.g., NA62, NA64) experiments.

In order to illustrate the phenomenological effect of a scenario which
involves sterile fermions, we focus on a minimal model which extends the Standard Model with an arbitrary number of
sterile states with masses in the above-mentioned ranges, known as
$3+N$ models, without making assumptions concerning the neutrino mass
(and/or the lepton mixing) generation mechanism, but allowing one to
access the degrees of freedom of the sterile neutrinos (their masses,
their mixing with the active neutrinos and the new \textit{CP}-violating
phases).  In this work, we address the phenomenological imprints of
such sterile fermions on observables involving kaon meson rare decays
into lepton final states including neutrinos, which can either be used 
to set (or {\it update}) constraints to model building, or provide
interesting observables that could be used as tests of various
scenarios of new physics. Notice that the first study devoted to probe massive neutrinos and lepton mixing using leptonic pseudoscalar
light meson decays was done in Refs.~\cite{Shrock:1980vy,Shrock:1980ct}.  These rare and forbidden decays are being
searched for at CERN (NA62) in the charged decay
modes~\cite{Moulson:2013grr} and at the J-PARC facility (KOTO)~\cite{Ahn:2016kja} and at CERN (NA64)~\cite{NA64,Gninenko:2016rjm} in the
neutral ones.  This study might also be useful for the TREK/E36 
experiment at J-PARC, where the data analysis is currently under way~\cite{Bianchin:2016vds}. It will further test the lepton universality 
in kaon two-body decays ($K_{\ell2}$) and search for a heavy neutrino~\cite{Kohl:2013rma}. Finally, this study could also be of use to
the proposed SHIP experiment~\cite{Alekhin:2015byh,Anelli:2015pba}, where we propose to use the large number of kaons produced in $D$-meson decays
to search for forbidden decays.

Having sterile neutrinos that are sufficiently light to be produced 
with non-negligible active-sterile mixing angles may induce important
impact on electroweak precision and many other observables.  Our
analysis must therefore comply with abundant direct and indirect
searches that have already allowed to put constraints or bounds on the sterile
neutrino masses and their mixing with the active neutrinos.

 \bigskip 
 In this work, we assume the effective case in which the Standard
 Model is extended by one sterile fermion (3 +1 case) and revisit weak
 kaon decays such as the leptonic $K_{\ell 2}$ ($K\to\ell{\nu}$), as
 well as the semileptonic ones $K_{\ell3}$ ($K\to \pi
 \ell{\nu}$). Furthermore we consider the loop-induced weak decays
 $K\to \pi\nu\nu$ and $K_L\to \nu\nu$ and derive analytical expressions
 for their decay rates, which are new results. In doing the numerical
 analysis, we (re)derive the expressions for various processes which
 are used as constraints ($\mu\to e\gamma$, $\mu\to eee$, $\tau\to
 \ell\nu\nu$, \dots).  We have chosen to present the
 analytical formulas and numerical results assuming here neutrinos to
 be Majorana fermions. A detailed discussion and comparison between
 the Dirac and Majorana cases is also displayed (in
 Appendix~\ref{app:a2}).
\bigskip 

 Our study reveals that the influence of the
 presence of massive sterile neutrinos on the $K_{\ell 2}$, $K_{\ell 3}$ and $K\to \pi \nu\nu$
 decay rates is less than $1\%$, thus fully compatible with the
 Standard Model predictions. Interestingly, however, we find that $\cb
 (K_L\to \nu\nu)$, which is zero in the Standard Model, can be as high
 as $\mathcal{O}( 10^{-10})$ and thus {possibly} within  
the reach of the NA62(-KLEVER), SHIP as well as the KOTO
 experiments. In other words, if all the future measurements of weak
 kaon decays turn out to be compatible with the Standard Model, the
 paradigm of the existence of sterile neutrinos would still remain
 valid. Instead, we show that  sterile neutrinos with mass below
 $m_K$ could generate a huge enhancement of $\cb (K_L\to \nu\nu)$.

\bigskip 
Our work is organized as follows: in Sec.~\ref{sec:sterile} we present
a generic model with one sterile neutrino added to the Standard Model
and present details concerning the parametrization used in the ensuing
analysis.  Sec.~\ref{sec:scan} is devoted to a discussion of
quantities which are used to constrain the parameter space followed by
the actual scan.  The expressions for various weak decay processes of
kaons are scrutinized in Sec.~\ref{sec:kaons}, and the
sensitivity of these processes on the presence of massive sterile
neutrinos is examined and discussed in Sec.~\ref{sec:discussion}.  Our
concluding remarks are given in Sec.~\ref{sec:conc}, while the Feynman
rules for the case of Majorana neutrino have been relegated to
Appendix~\ref{app:a1}. Appendix~\ref{app:a2} contains a comparison of 
the analytical expressions for $\cb (K\to \pi \nu\nu)$ when neutrinos are 
Majorana or Dirac fermions.

\section{Extending the Standard Model with Sterile Fermions}
\label{sec:sterile}
\subsection{Models with sterile fermions}
To discuss the phenomenological consequences of sterile fermion states on
 low-energy physics observables, it is important to have an idea of
the underlying framework which involves sterile neutrinos, and among
those the testable ones in particular. Such models are for instance
 based on low-scale seesaw mechanism, e.g. extension of the
Standard Model by {\it exclusively} right-handed (RH) neutrinos, like
in the usual type~I seesaw, which is realized at the TeV
scale~\cite{Minkowski:1977sc,Ramond:1979py,Yanagida:1979as,GellMann:1980vs,Glashow:1979nm,Mohapatra:1979ia,
  Schechter:1980gr,Schechter:1981cv}, and for which the Yukawa
couplings are small ($\sim Y_e$). They can nevertheless be made higher
if one assumes some extra input like the minimal flavor
violation~\cite{Gavela:2009cd,Ibarra:2010xw}), or tiny as it is the
case in $\nu$MSM~\cite{Asaka:2005an}. Other than RH neutrinos,
$\nu_R$, one can also consider {\it additional sterile fermions} with
the opposite lepton numbers of the RH ones, like it is done in the
case of the {\it linear}~\cite{Barr:2003nn,Malinsky:2005bi} or the
{\it
  inverse}~\cite{Mohapatra:1986bd,GonzalezGarcia:1988rw,Deppisch:2004fa,Abada:2014vea}
seesaw mechanisms. The two latter scenarios are (theoretically and
phenomenologically) very appealing as they provide an extra suppression
factor, which is linked to a small violation of the total lepton
number, allowing one to explain the tininess  of neutrino masses while
having large Yukawa couplings and a comparatively low seesaw
scale. \\ Having relatively light sterile fermions which do not
decouple, since they can have non-negligible active-sterile mixing,
certainly leads to important consequences and  as a result to numerous
constraints.  The most important and direct consequence is the
modification of the charged and neutral currents as
\bea
\label{mod.current} \mathcal{L}_{W^\pm} &\supset&
-\frac{g_2}{\sqrt{2}} \, W^-_\mu \, \sum_{\alpha=e,\mu,\tau}
\operatornamewithlimits{\sum}_{i=1}^{3 + N} U_{\alpha i} \, \bar
\ell_\alpha\, \gamma^\mu\, P_L\, \nu_i\ , \\\nonumber
\mathcal{L}_{Z}&\supset& -\frac{g_2}{4 \cos \theta_W} \, Z_\mu \,
\sum_{i,j=1}^{3 + N} \bar \nu_i \,\gamma ^\mu \left[ P_L\,( U^\dagger
  U)_{ij} - P_R \, ( U^\dagger U)_{ij}^* \right] \nu_j\ , 
  \eea 
where $g_2$ is the weak coupling constant, and $N$ is the number of sterile fermions. 
The modified lepton mixing matrix, obviously nonunitary, also encodes the active-sterile mixing. In the limit in which the
sterile fermions decouple the matrix $U$ corresponds to the usual
Pontecorvo--Maki--Nakagawa--Sakata $3\times 3$ unitary matrix, i.e., $
U_{\alpha i}= U_\text{PMNS}$.\\ Moreover, if sufficiently light, the
sterile neutrinos can be produced as decay products. Both these points
might induce a huge impact on numerous observables, which in turn can
provide abundant constraints on the sterile fermions (masses and
the active-sterile mixing angles including the new \textit{CP}-violating phases).
  
A useful first approach to study the impact of sterile neutrinos on
the low-energy processes relies on addition of only one sterile fermion
to the Standard Model with no hypothesis regarding the origin of the
light neutrino masses and the observed lepton mixing
($U_\text{PMNS}$).

\subsection{Effective Approach: Standard Model + one sterile fermion}
\label{sec:eff}
In essence, since no seesaw hypothesis is made, the physical
parameters correspond to the three mostly active neutrino masses, the
mass of the mostly sterile neutrino, and finally the mixing angles and
the \textit{CP}-violating phases encoded in the mixing matrix which relates the
physical neutrino to the weak interaction basis.  Due to the modification of the charged
current in \eqref{mod.current} the lepton mixing matrix is defined as
\bea
\label{Udef}
U_{\alpha i}\, =\,\sum_{k=1}^3 V^*_{k\alpha}\, U_{\nu_{ki}}\, ,
\eea
where, $V$ and $U_\nu$ are the unitary transformations that relate the physical charged and neutral lepton
 states $\ell$ and $\nu$ to the gauge eigenstates $\ell'$ and $\nu'$ as
\bea
\label{RotDiag}
\ell'_L =\, V\, \ell_L \,,\qquad  \nu'_L \,=\, U_\nu\, \nu_L\, .
\eea

In the $3+1$ model, the mixing matrix $U$ includes six rotation angles,
three Dirac \textit{CP}-violating phases, in addition to the three Majorana phases. It can thus be parametrized as 
 follows: 
\bea\label{eq:para1}
 U^T &=& R_{34}(\theta_{34},\delta_{43}) \cdot R_{24}(\theta_{24}) \cdot R_{14}(\theta_{14},\delta_{41}) 
\cdot \tilde{U} \cdot \rm diag(\phi_{21},\phi_{31},\phi_{41})\ ,
\eea
where $R_{ij}$ is the rotation matrix between $i$ and $j$, which includes the mixing angle $\theta_{ij}$ and the Dirac \textit{CP}-violating phase $\delta_{ij}$. The Majorana \textit{CP}-violating phases are factorized in the last term of Eq.~(\ref{eq:para1}), where 
$\phi_{ij}= \exp^{-i\left(\phi_i-\phi_j\right)}$. $\tilde{U}$ is the  $4\times 3$ matrix which encodes the mixing among the active leptons as
\bea
\tilde{U} = 
\left(
\begin{array}{ccc}
 U_{e1} &
 U_{e2} &
 U_{e3} 
 \\ 
 {U_{\mu 1}} &
 {U_{\mu 2}} &
 {U_{\mu 3}} 
 \\
 {U_{\tau 1}} &
 {U_{\tau 2}} &
 {U_{\tau 3}} 
\\
0&0&0 
\end{array}
\right)\ . 
\eea
The upper $3\times 3$ submatrix of $\tilde{U}$  
is  nonunitary due to the presence of a sterile neutrino and  includes the usual Dirac \textit{CP} phase actively
searched for in neutrino oscillation facilities. In the case where the sterile neutrino decouples, this submatrix would correspond to
the usual unitary PMNS lepton mixing matrix, ${U}_\text{PMNS}$.  The active-sterile mixing is described by the rotation matrices $R_{34},R_{24},R_{14}$ which are defined as
\begin{eqnarray}  
R_{34}\ &=&\ \left( 
\begin{array}{cccc}
1 & 0 & 0 & 0 \\
0 & 1 & 0 & 0 \\
 0 & 0 & \rm cos \theta_{34} &\rm sin \theta_{34} \cdot e^{-i \delta_{43}}\\ 
 0 & 0 & -\rm sin \theta_{34} \cdot e^{i \delta_{43}}& \rm cos \theta_{34} 
\end{array}%
\right) \ ,\nonumber
\end{eqnarray} 
\begin{eqnarray} \label{eq:R}
R_{24}\ &=&\  \left( 
\begin{array}{cccc}
1 & 0 & 0 & 0 \\
0 & \rm cos \theta_{24}  & 0 & \rm sin \theta_{24}\\ 
0 & 0 & 1 & 0 \\
0 & - \rm sin \theta_{24}& 0 &  \rm cos \theta_{24} %
\end{array}%
\right)\ , \nonumber \\
\hfill \nn \\
\hfill \nn \\
R_{14}\ &=& \left( 
\begin{array}{cccc}
\rm cos \theta_{14} & 0 & 0 & \rm sin \theta_{14} \cdot e^{-i \delta_{41}} \\
0 & 1 & 0 & 0 \\
0 & 0 & 1 & 0 \\
- \rm sin \theta_{14} \cdot e^{i \delta_{41}} & 0 & 0 &\rm cos \theta_{14} \\
\end{array}%
\right) \ .
\end{eqnarray}

\section{Scan of the parameter space in the $3+1$ effective approach\label{sec:scan}} 

In this section we list the quantities that are used in order to
constrain the parameter space of the scenario with three active and
one (effective) sterile neutrino.~\footnote{The word effective is used to denote the fact that this sterile neutrino is 
mimicking the effect of several fermionic singlets usually induced when embedding the Standard Model with a fermion-type seesaw.}  In addition to the current limits on
the neutrino data~\cite{Esteban:2016qun}, the presence of an extra sterile neutrino requires the introduction of new parameters: its mass, 
three new (active-sterile) mixing angles
and two extra \textit{CP}-violating phases. Furthermore,  since we assume
that neutrinos are Majorana fermions, there are also three Majorana
phases which, however, do not play a significant role in the setup
discussed in this paper.
  
Before we list the observables used to constrain
the parameter space, we need to emphasize that a price to pay for
adding massive sterile neutrinos is that the Fermi constant -- extracted from the muon decay -- should be
redefined according to $G_F=G_\mu/\sqrt{ \sum_{i,j} |U_{ei}|^2 |U_{\mu
    j}|^2}$, where the sum runs over kinematically accessible neutrinos. We checked, however, that for the model used in this
paper $G_F = G_\mu$ remains an excellent approximation and thus it will be
used in the following. \\

\noindent
$\bullet$ \underline{$\mu \to e\gamma$}: An important constraint comes from the combination of the recently
established experimental bound $\cb(\mu\to e\gamma) < 4.2\times
10^{-13}$~\cite{Adam:2013mnn}. In our setup, the branching fraction of this decay is given by the following
expression~\cite{Ilakovac:1994kj}:
\begin{align}
\cb (\mu \to e\gamma) &= {\sqrt{2} G_F^3 s_W^2 m_W^2 \over 128 \pi^5 \Gamma_\mu} m_\mu^5 \left\vert \sum_{i=1}^4 U_{\mu i}^\ast U_{e i} G_\gamma (x_i)\right\vert^2\,, \cr
G_\gamma (x) & = - { 2 x^3+5x^2 -x\over 4(1-x)^3 }
- {  3 x^3  \over 2(1-x)^4 } \log x\ ,
\end{align} 
where $x_i= m_{\nu_i}^2/m_W^2$ and $s_W^2=1-m_W^2/m_Z^2$. We also use the above expression, {\sl mutatis mutandis}, to derive additional constraints stemming from  
the experimental limits, $\cb(\tau\to \mu\gamma)<4.4\times 10^{-8}$ and $\cb(\tau\to e\gamma)<3.3\times 10^{-8}$~\cite{PDG}. \\

\noindent $\bullet$ \underline{$W \to \ell \nu$}: Combining the measured $\cb(W\to e\nu )=0.1071(16)$ and $\cb(W\to \mu\nu)=0.1063(15)$, with the expression
\begin{align}
\cb(W \to \ell  \nu) = {\sqrt{2} G_F m_W \over 24 \pi  \Gamma_W} \sum_{i=1}^4 \lambda^{1/2}(m_\ell^2,m_{\nu_i}^2,m_W^2) \left[ 
2 - { m_\ell^2+m_{\nu_i}^2\over m_W^2}- { (m_\ell^2-m_{\nu_i}^2)^2\over m_W^4} 
\right] \vert U_{\ell i}^2\vert  ,
\end{align} 
yields useful constraints in the parameter space. In the above formula
$\lambda(a^2,b^2,c^2)= [a^2 - (b+c)^2]\, [a^2 - (b-c)^2]$.  Since we
do not include the electroweak radiative corrections to this formula
we will use in our scan the experimental results with $3\sigma$
uncertainties.  Notice also that unlike $\cb(W\to e\nu )$ and
$\cb(W\to \mu\nu)$, which have also been recently measured at the 
LHC~\cite{Aaij:2016qqz}, the LEP result for $\cb(W\to \tau\nu )$ has
not been measured at the LHC.  For that reason, and despite the fact that
the LEP result for $\cb(W\to \tau\nu )$ differs from the Standard
Model value at the $2.3\sigma$ level, we prefer not to include
$\cb(W\to \tau\nu )$ in our scan. \\

\noindent $\bullet$ \underline{$\Delta r_{\pi}= r_{\pi}^{\rm
    exp}/r_{\pi}^{\rm SM} - 1$}: The ratio $r_\pi=\Gamma(\pi\to e \nu_e
)/\Gamma(\pi\to \mu \nu_\mu )$ provides an efficient constraint, as
recently argued in Ref.~\cite{Abada:2013aba}.  To that end one
combines the Standard Model expression for the decay rate with the
experimental values to obtain $\Delta r_{\pi}= 0.004(4)$, the result
which is then compared with the formula relevant to the scenario
discussed in this paper (cf. next section), namely,
\begin{align}\label{rp}
\Delta r_\pi= - 1 + {m_\mu^2 (m_\pi^2-m_\mu^2)^2\over m_e^2 (m_\pi^2-m_e^2)^2} {\displaystyle{\sum_{i=1}^4} \vert U_{ei}\vert^2 \left[ m_\pi^2 (m_{\nu_i}^2 + m_e^2) -  (m_{\nu_i}^2 - m_e^2)^2 \right] \lambda^{1/2}(m_\pi^2,m_{\nu_i}^2 ,m_e^2) 
\over \displaystyle{\sum_{i=1}^4} \vert U_{\mu i}\vert^2 \left[ m_\pi^2 (m_{\nu_i}^2 + m_\mu^2) -  (m_{\nu_i}^2 - m_\mu^2)^2 \right] \lambda^{1/2}(m_\pi^2,m_{\nu_i}^2 ,m_\mu^2)
}.
\end{align} 
In this way one gains another interesting constraint to the parameter space.\\

\noindent $\bullet$ \underline{$Z \to \nu \nu$}: In addition to the active neutrinos, the sterile ones can be used to saturate the experimental  $Z$ invisible decay width, ${\Gamma}(Z\to \text{invisible} )=0.503(16)$~GeV~\cite{PDG}. The corresponding expression, 
which we compute by using the Feynman rules derived in Appendix~\ref{app:a1}, reads~\footnote{We reiterate that all along this paper we consider neutrinos to be Majorana fermions.}
\begin{align}
\Gamma(Z \to \nu \nu) =&\mathop{ \sum_{i,j=1}^{4}}_{i\leq j}  \left(1-\frac{\delta_{ij}}{2}\right)  {\sqrt{2} G_F \over 24 \pi }m_Z \lambda^{1/2}(m_Z^2,m_{\nu_i}^2,m_{\nu_j}^2)  \nn\\
                                              &\times  \left[  |C_{ij}|^2 \left( 2 -\frac{m_{\nu_i}^2 + m_{\nu_j}^2}{m_Z^2} - \frac{(m_{\nu_i}^2 - m_{\nu_j}^2)^2}{m_Z^4} \right) - 
                                              \mathrm{Re}\left( C_{ij}^2\right) { 6m_{\nu_i}m_{\nu_j} \over m_Z^2}\right]\,,
\end{align} 
where
\bea
C_{ij} = \sum_{\alpha\in \{e,\mu,\tau\}} U_{\alpha i}^\ast U_{\alpha j}\,.
\eea
\\

\noindent $\bullet$ \underline{$\ell^\prime \to \ell\ell\ell$}: To implement this constraint we compare the experimental limit, $\cb(\mu \to e e e) <  1 \times 10^{-12}$~\cite{Bellgardt:1987du}, to the theoretical prediction derived in Ref.~\cite{Ilakovac:1994kj}:
\bea
\cb (\mu \to eee) &=& \frac{G_F^4 m_W^4 }{6144 \pi^7}\frac{m^5_\mu}{\Gamma_\mu}\nn\\ 
&&  \left\{ 2 \left|\frac{1}{2}F^{\mu eee}_{\rm Box}+F^{\mu e}_Z-2\sin^2\theta_W (F^{\mu e}_Z-F^{\mu e}_\gamma)\right|^2+4 \sin^4\theta_W \left|F^{\mu e}_Z-F^{\mu e}_\gamma\right|^2 \right. \nn\\ 
&& \left.		+ 16 \sin^2\theta_W \mathrm{Re} \left[	(F^{\mu e}_Z +\frac{1}{2}F^{\mu eee}_{\rm Box})	{G^{\mu e}_\gamma}^* 			\right]		- 48 \sin^4\theta_W \mathrm{Re}\left[	(F^{\mu e}_Z-F^{\mu e}_\gamma)	{G^{\mu e}_\gamma}^* 			\right] \right. \nn\\ 
&& \left.
		+32 \sin^4\theta_W |G^{\mu e}_\gamma|^2\left[		\ln \frac{m^2_\mu}{m^2_{e}} -\frac{11}{4}		\right]		\right\},
\eea
where the explicit forms of the loop functions $F^{\mu eee}_{\rm Box},F^{\mu e}_Z,F^{\mu e}_\gamma,G^{\mu e}_\gamma$ can be found in Refs.~\cite{Ilakovac:1994kj,Alonso:2012ji}.
Similarly, we implement in our scan the bounds arising from the experimental limits $\cb(\tau \to \mu\mu\mu) < 2.1\times  10^{-8}$, and $\cb(\tau \to e e e) < 2.7\times  10^{-8}$~\cite{PDG}.\\

\noindent $\bullet$ The leptonic decays $\tau \to \ell \nu\nu$ ($\ell = e,\mu$) represent very useful constraints as well. We derived the relevant expression for this process and found
\begin{align}
\frac{d\cb(\tau \to \ell \nu\nu)}{dq^2} =& \mathop{ \sum_{i,j=1}^{4}}_{i\leq j}  \left(1-\frac{\delta_{ij}}{2}\right)  { G_F^2 \tau_\tau \over 192 \pi^3 m_\tau^3 q^6 } \lambda^{1/2}(m_\tau^2,m_\mu^2,q^2)\lambda^{1/2}(q^2m_{\nu_i}^2,m_{\nu_j}^2) \Biggl\{ \left( |U_{\tau i} U_{\ell j}^\ast|^2 + |U_{\tau j} U_{\ell i}^\ast|^2 \right)
\Biggr.\nn\\
&  \times \biggl[ 3 \biggl(q^4 - (m_{\nu_i}^2-m_{\nu_j}^2)^2 \biggr) 
 \biggl( (m_\tau^2-m_\ell^2)^2 - q^4\biggr) - \lambda(m_\tau^2,m_\mu^2,q^2)\lambda(q^2m_{\nu_i}^2,m_{\nu_j}^2) \biggr] \nn\\
&\Biggl. - 24 \; \mathrm{Re}\left( U_{\tau i}^\ast U_{\ell j} U_{\tau j} U_{\ell i}^\ast\right) m_{\nu_i}m_{\nu_j} q^4 (m_\tau^2 +m_\ell^2 - q^2) \Biggr\}.
\end{align}
The above formula is then combined with the average of experimental results summarized in Ref.~\cite{PDG}, namely $\cb (\tau\to \mu\nu\nu)=17.33(5)\%$, and $\cb (\tau\to e\nu\nu)=17.82(5)\%$. 

 \bigskip 

\begin{figure}[ht!]
\centering
\includegraphics[width=0.65\linewidth]{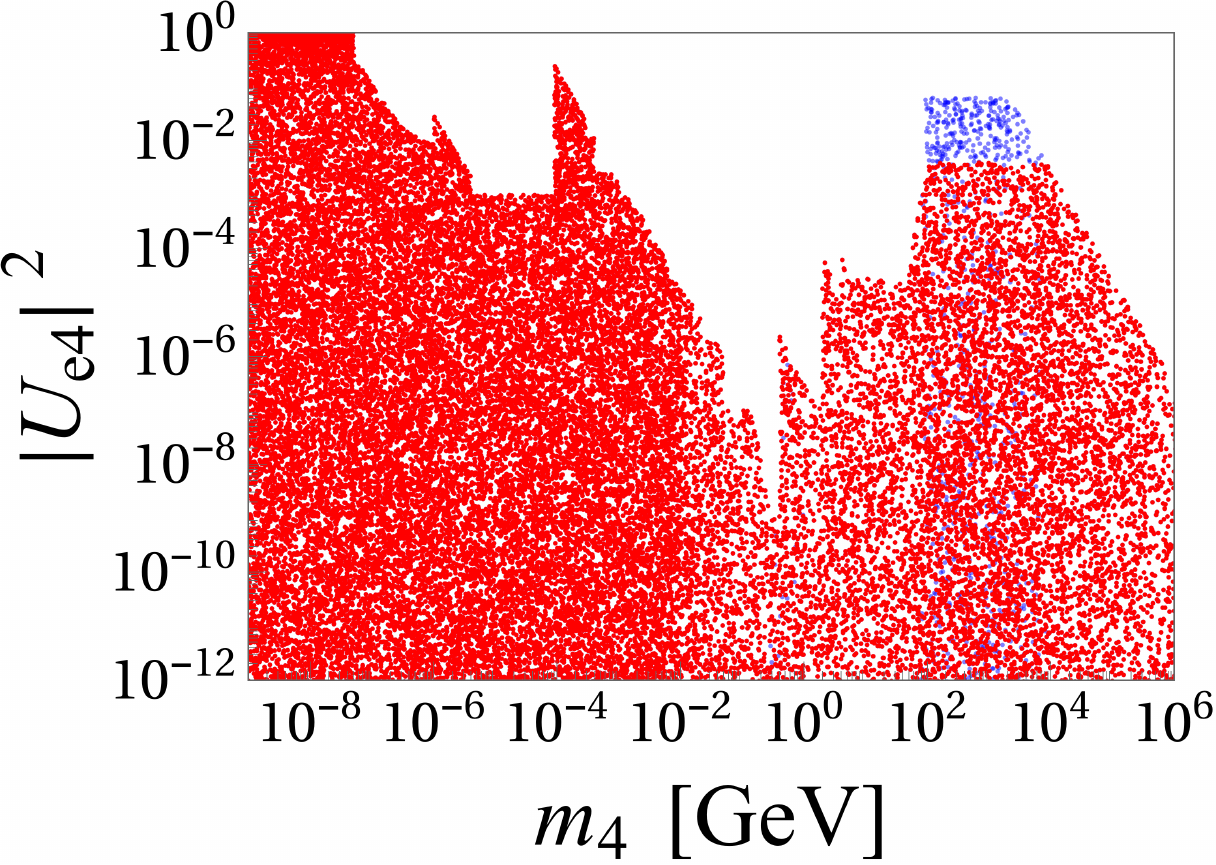}
\caption{\small \sl Result of the scan in the scenario
  of three active and one (effective) sterile neutrino, displayed in the
  plane $|U_{e4}|^2$ vs $m_4$. Perturbative unitarity cuts the
  parameter space for large $m_4$. The red points agree with all constraints while the blue ones are
  excluded by requiring the compatibility with
  experimental results for the leptonic $\tau$ decays.}
\label{fig:1}
\end{figure}

\begin{figure}[ht!]
\centering
\includegraphics[width=0.33\linewidth]{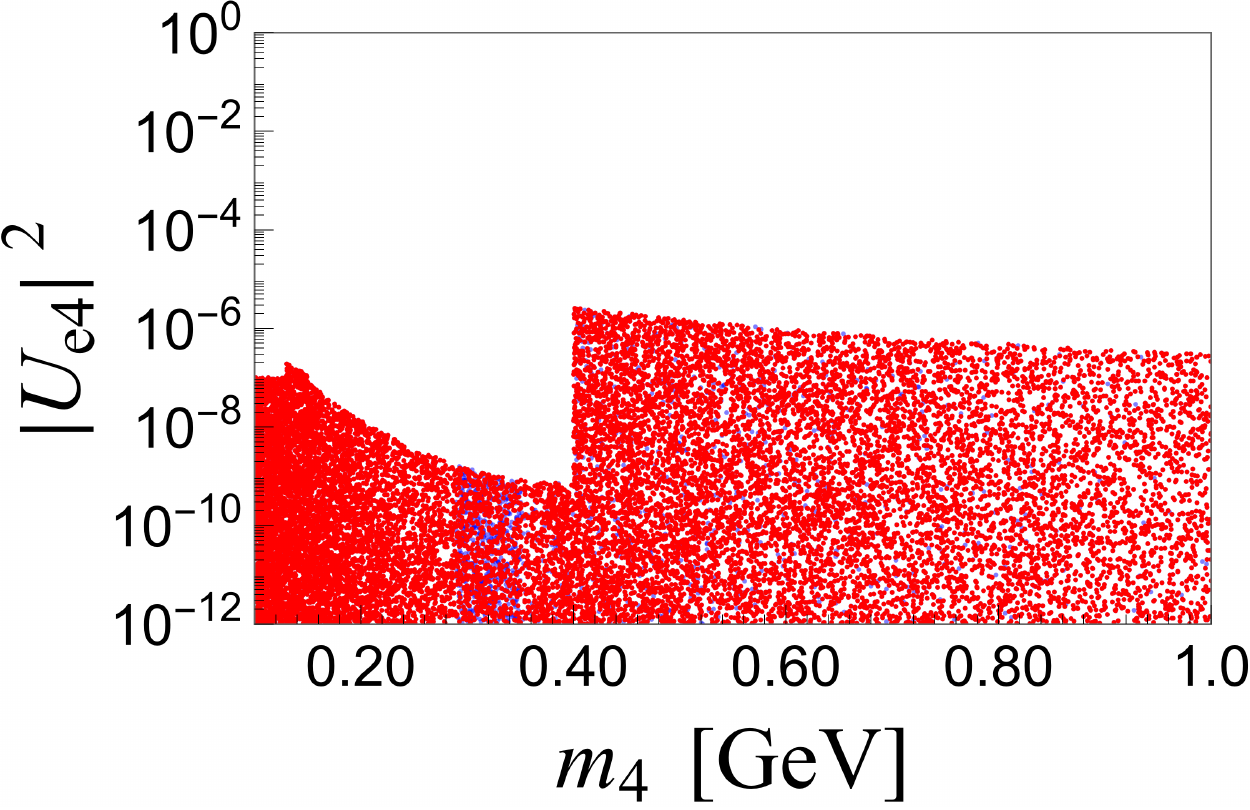}~\includegraphics[width=0.33\linewidth]{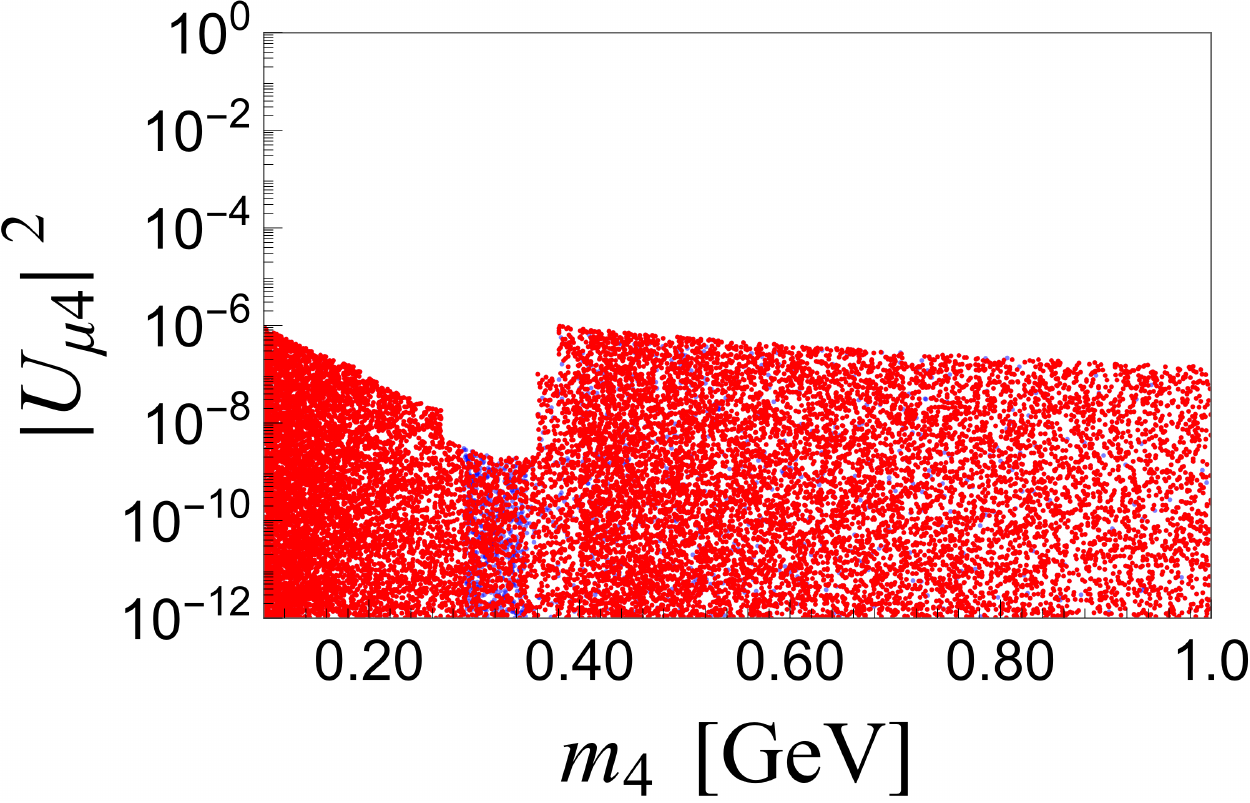}~\includegraphics[width=0.33\linewidth]{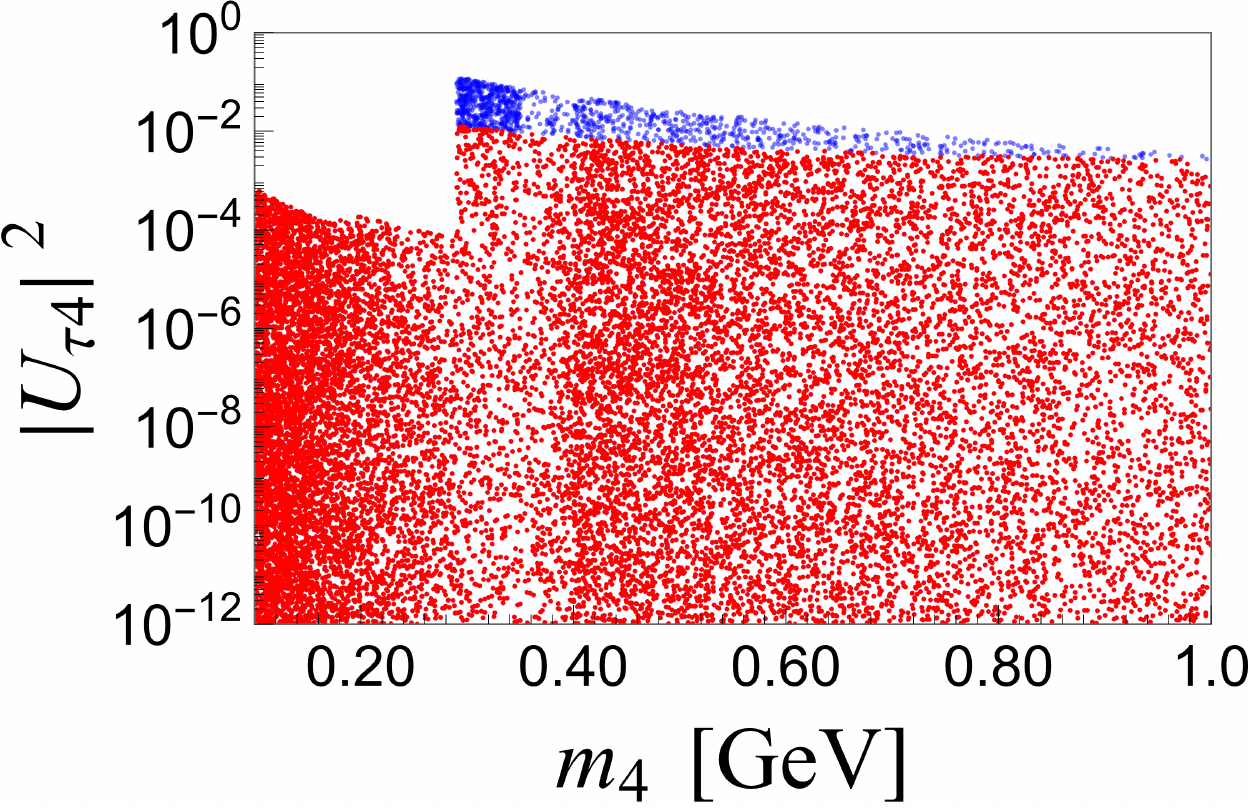}
\caption{\small \sl As in Fig.~\ref{fig:1} but for $m_4\leq 1~\gev$, and for all $|U_{\ell 4}|^2$. }
\label{fig:2}
\end{figure}

 We perform a first random scan of $100 000$ points using flat priors on the Dirac \textit{CP} phases and logarithmic priors on all other scan parameters, which are chosen in the following
ranges
\begin{align}
 10^{-21}\,\mathrm{eV} \leq\, 	& m_{\nu_1} \leq 1\,\mathrm{eV}\,, \nonumber \\
 10^{-9} \,\mathrm{GeV} \leq\, 	& m_{\nu_4} \leq 10^6\,\mathrm{GeV}\,, \nonumber \\
 10^{-8} \leq \theta_{14},\, 	& \theta_{24},\, \theta_{34} \leq 2 \pi\,, \\
 0 \leq \delta_{13},\, 		& \delta_{41},\, \delta_{43} \leq 2 \pi\,. \nonumber
\end{align}
We then perform a second random focusing on the window where the heavy neutrino mass is comparable to the kaon mass, using flat priors on the Dirac \textit{CP} phases
and logarithmic priors on all other scan parameters. We first generate a sample of $200 000$ points with parameters chosen as
\begin{align}
 10^{-21}\,\mathrm{eV} \leq\, 	& m_{\nu_1} \leq 1\,\mathrm{eV}\,, \nonumber \\
 0.1 \,\mathrm{GeV} \leq\, 			& m_{\nu_4} \leq 1\,\mathrm{GeV}\,, \nonumber \\
 10^{-6} \leq \theta_{14},\,	& \theta_{24},\, \theta_{34} \leq 2 \pi\,, \\
 0 \leq \delta_{13},\, 		& \delta_{41},\, \delta_{43} \leq 2 \pi\,, \nonumber
\end{align}
to which we add $40 000$ points with paramaters chosen in the ranges
\begin{align}
 10^{-21}\,\mathrm{eV} \leq\, 	& m_{\nu_1} \leq 1\,\mathrm{eV}\,, \nonumber \\
 0.27 \,\mathrm{GeV} \leq\, 	& m_{\nu_4} \leq 0.35\,\mathrm{GeV}\,, \nonumber \\
 10^{-6} \leq\,			& \theta_{14},\,\theta_{24} \leq 2 \pi\,, \\
 0.1 \leq 			& \theta_{34} \leq 2 \pi\,, \nonumber \\
 0 \leq \delta_{13},\, 		& \delta_{41},\, \delta_{43} \leq 2 \pi\,. \nonumber
\end{align}
The other parameters are fixed from the best fit point in~\cite{Esteban:2016qun},i.e.
\begin{align}
\sin^2 \theta_{12}\,=\, 0.306\,, 
\quad
\sin^2 \theta_{23}\,=\, 0.441\,, 
\quad
\sin^2 \theta_{13}\,=\, 0.02166\,, \nonumber \\
\Delta m^2_{21}\,=\, 7.50\times 10^{-5} \mathrm{eV}^2\,, 
\quad
\Delta m^2_{31}\,=\, 2.524\times 10^{-3} \mathrm{eV}^2\, . 
\end{align}
We then impose all of the above constraints, in
addition to those arising from the direct searches~\cite{Atre:2009rg},
and require the perturbative unitarity
condition~\cite{Ilakovac2} which can be written as
\bea
\label{pert-uni} {G_F m_4^2 \over \sqrt{2}\pi}\sum_\alpha \vert
U_{\alpha 4}\vert^2 < 1\,.  
\eea 
That last condition is important when
the sterile neutrino is very heavy as it leads to its decoupling,
which can be seen in Fig.~\ref{fig:1}.\footnote{Notice that we often
  use the notation $m_{\nu_4}\equiv m_4$ which should not be confusing
  to the reader.}
For the purpose of this paper, in which we study the effects of an
additional sterile neutrino on the kaon physics observables, the most
interesting region is the one corresponding to $m_4\lesssim 1\;\gev$,
which we show in Fig.~\ref{fig:2}. As expected, the limits coming
from the $\tau$ leptonic decays are the most constraining in the plane
$(m_4, |U_{\tau 4}|^2)$. Notice also that the sharp exclusion of
parameters around $m_4 \sim 0.3\ \gev$ comes from the direct searches
discussed in Ref.~\cite{Atre:2009rg}. {It is worth noticing that the bounds shown in Fig.~\ref{fig:2} are in agreement with those provided in Ref.~\cite{Atre:2009rg} in the considered mass regime, although  slightly improved as most of the constraints discussed above have been updated.}

In summary, we selected the points in the parameter space which are
compatible with a number of constraints discussed in the body of this
section.  We will use the results of the above scan to test the
sensitivity of the kaon physics observables on the presence of an {\sl
  effective} massive sterile neutrino with a mass $m_4\lesssim 1$~GeV.
\section{Kaon Physics Phenomenology}
\label{sec:kaons}
Before discussing the results stemming from our scan, we will introduce the kaon decays 
at the heart of our study and present their analytical expressions in our effective model. In our phenomenological discussion, we will consider the processes for which hadronic uncertainties are under full theoretical control by means of numerical simulations of QCD on the lattice. Processes such as $\mathcal{B}(K_S\to \mu\mu)$ and $\mathcal{B}(K\to \pi\mu\mu)$ will not be considered, since the corresponding Standard Model predictions depend on large (long-distance QCD) uncertainties.

\subsection{Leptonic decays, $K_{\ell 2}$}
Sterile neutrinos in kaon and pion leptonic decays were first studied and analyzed in~\cite{Shrock:1980vy} with the aim to probe massive neutrinos via lepton mixing; correspondingly, associated tests  allowed one to set bounds on neutrino masses and lepton mixing matrix elements~\cite{Shrock:1980ct}. Here we revisit the $K_{\ell 2}$ decays in light of the existing data on neutrinos and in the framework of the simple extension of the Standard Model by one sterile fermion with the aim to update the latter obtained results. 
 
The effective Hamiltonian we will be working with reads
\bea
\label{Heff:1}
\mathcal{H}_\eff = \sqrt{2} G_F V_{us} \left[ (1+g_V)\, \bar u
  \gamma_\mu s\, \bar \ell_L \gamma^\mu \nu_L - (1+g_A)\, \bar u
  \gamma_\mu \gamma_5 s\, \bar \ell_L \gamma^\mu \nu_L +
  \mathrm{h.c.}\right]\,, \eea where $g_A$ and $g_V$ are the generic
couplings to physics beyond the Standard Model, which in our case are
the couplings to the massive sterile neutrino. {The leptonic bilinear  $\bar \ell_L \gamma^\mu \nu_L$ in Eq.~(\ref{Heff:1}) should be understood as $\sum_{\alpha=e,\mu,\tau}
\operatornamewithlimits{\sum}_{i=1}^{3 + N} U_{\alpha i} \, \bar
\ell_\alpha\, \gamma^\mu\, P_L\, \nu_i\ $, as in Eq.~(\ref{mod.current}). In the effective approach, the effect of the active-sterile mixing is encoded in the effective couplings $g_A$ and $g_V$}.  The relevant hadronic
matrix element for this decay is parametrized in terms of the decay
constant $f_K$ via \bea\label{eq:fk} \langle 0\vert \bar{u}\gamma_\mu
\gamma_5 s\vert K^-(p)\rangle = i f_Kp_\mu, \eea so that the decay
amplitude becomes \bea \mathcal{A}= -i \sqrt{2} G_FV_{us} (1+g_A) (i
f_K p_\mu) \bar u(k_1)\gamma^\mu P_Lv(k_2) \,, \eea where $k_1$ and
$k_2$ are the momenta of the lepton and neutrino, respectively, and
$P_L=(1-\gamma_5)/2$. Multiplying this amplitude by its conjugate
and after summing over the spins we then get \bea \sum_\mathrm{spin}
|\mathcal{A}|^2 = 2 G_F^2 |V_{us}|^2 f_K^2 |1+g_A|^2 \left[ m_K^2
  (m_\ell^2 +m_\nu^2) - (m_\ell^2 -m_\nu^2)^2\right], \eea so that the
final expression for the decay rate reads
\begin{align}
\cb (K\to \ell \nu_i)={G_F^2\tau_K\over 8\pi m_K^3}  |V_{us}|^2 f_K^2 \lambda^{1/2}(m_K^2,m_\ell^2,m_{\nu_i}^2) |1+g_A|^2 \left[ m_K^2 (m_\ell^2 +m_{\nu_i}^2) - (m_\ell^2 -m_{\nu_i}^2)^2\right] \,.
\end{align}
{One can immediately see from the above equation (and the subsequent ones for the observables under this study)  that the presence of the sterile state can have two consequences, a phase space effect if its mass is kinematically allowed and  a modification of the coupling due to the active-sterile mixing encoded in $g_A$. }
More explicitly, and after adapting the above formula to the scenario with an extra sterile neutrino, we have
\begin{align}\label{eq:BrKl2}
\cb (K\to \ell \nu )={G_F^2\tau_K\over 8\pi m_K^3}  |V_{us}|^2 f_K^2  
\sum_{i=1}^4 |U_{\ell i}|^2 \lambda^{1/2}(m_K^2,m_\ell^2,m_{\nu_i}^2)   \left[ m_K^2 (m_\ell^2 +m_{\nu_i}^2) - (m_\ell^2 -m_{\nu_i}^2)^2\right] \,.
\end{align}
Similarly, for the process $\tau \to K\nu$ we get
\begin{align}
\cb (\tau\to K  \nu)={G_F^2 \tau_\tau \over 16\pi m_\tau^3} |V_{us}|^2 f_K^2\sum_{i=1}^4 |U_{\tau i}|^2   \lambda^{1/2}(m_\tau^2,m_K^2,m_{\nu_i}^2)   \left[  (m_\tau^2 -m_{\nu_i}^2)^2 - m_K^2 (m_\tau^2 +m_{\nu_i}^2) \right] \,.
\end{align}
The above expressions can be trivially extended to the case of the
pion leptonic decay by simply replacing $K\to \pi$, and $V_{us}\to
V_{ud}$.  Modern day lattice QCD computations of the decay constants
$f_K$, $f_\pi$, and especially of $f_K/f_\pi$, have already reached a
subpercent accuracy~\cite{Aoki:2016frl} so that comparing
the theoretical expressions (in which the effects of new physics are
included) with the experimental measurements can result in stringent
constraints on the new physics couplings.

\subsection{Semileptonic decays, $K_{\ell 3}$}
To discuss the semileptonic decays $K\to \pi \ell\nu$, we again rely
on the effective Hamiltonian~\eqref{Heff:1} and keep the neutrinos
massive.  Due to parity, only the vector current contributes
on the hadronic side and the relevant hadronic matrix matrix element
is parametrized as 
\bea\label{eq:Kl3FF} \langle \pi^+(k)\vert \bar
u\gamma_\mu s\vert \bar{K}^0(p)\rangle = \left( k_\mu+k_\mu -
\frac{m_K^2-m_\pi^2}{q^2}q_\mu\right) f_+(q^2) +
\frac{m_K^2-m_\pi^2}{q^2}q_\mu \, f_0(q^2)\,, \eea 
where the form
factors $f_{+,0}(q^2)$ are functions of $q^2=(p-k)^2=(k_1+k_2)^2$
which can take the values $q^2\in [(m_\ell+m_\nu)^2, (m_K-m_\pi)^2]$.
Notice that the hadronic matrix element of the decay to a neutral pion is related to the above one by {isospin symmetry}, i.e. 
$\langle \pi^0(k)\vert \bar
u\gamma_\mu s\vert {K}^-(p)\rangle =(1/\sqrt{2})\ \langle \pi^+(k)\vert \bar
u\gamma_\mu s\vert \bar{K}^0(p)\rangle$. This decay is suitably described by its helicity amplitudes.  To that
end one first defines the polarization vectors of the virtual vector
boson ($K\to \pi V^\ast$) as 
\bea \varepsilon_\pm^\mu =
\frac{1}{\sqrt{q^2}} \left[ 0, \ \pm1, \ -i, \ 0\right]^T,\quad
\varepsilon_0^\mu = \frac{1}{\sqrt{q^2}} \left[ |\vec q|, \ 0, \ 0,
  \ -q^0\right]^T, \quad \varepsilon_t^\mu = \frac{1}{\sqrt{q^2}} \left[
  q^0, \ 0, \ 0, \ - |\vec q|\right]^T, \eea so that the only nonzero
helicity amplitudes will be $h_{0,t}(q^2) = (1+g_V) \varepsilon_{0,t}^{\mu\; \ast}
\langle \pi^+(k)\vert \bar u\gamma_\mu s\vert \bar{K}^0(p)\rangle$, or
explicitly 
\bea 
h_{0}(q^2) = {(1+g_V)\over \sqrt{q^2}} \lambda^{1/2}(m_K^2,q^2,m_\pi^2) f_+(q^2)\,,\quad h_{t}(q^2)
={(1+g_V)\over \sqrt{q^2}} \left(m_K^2-m_\pi^2 \right) f_0(q^2).  \eea
In terms of these functions the decay amplitude reads \bea
\mathcal{A}_{sl}= \sqrt{2} G_FV_{us} \left[ -h_0(q^2)
  \varepsilon_\mu^0 + h_t(q^2) \varepsilon_\mu^t \right]\; \bar
u(k_1)\gamma^\mu P_L v(k_2) \,.  \eea In the rest frame of the lepton
pair the components of the vectors $k_1$ and $k_2$ of the final
leptons are \bea k_1=(E_\ell, p_\ell \sin\theta, 0, p_\ell
\cos\theta), \quad k_2= (E_\nu, - p_\ell \sin\theta, 0, - p_\ell
\cos\theta ), \eea where \bea E_\ell = {q^2+m_\ell^2 - m_\nu^2\over
  2\sqrt{q^2}}, \quad E_\nu = {q^2-m_\ell^2 + m_\nu^2\over
  2\sqrt{q^2}} , \quad p_\ell = {\lambda^{1/2}(q^2,m_\ell^2,m_\nu^2)
  \over 2 \sqrt{q^2}}\,, \eea 
{and $\theta$ is the angle between $\ell^-$ (in the lepton pair rest frame)  and the flight direction of the leptonic pair (opposite to the pion direction) in the
kaon rest frame.} The decay rate can then be written as
\bea\label{eq:angular} 
\frac{d^2\cb (K\to \pi\ell{\nu})}{dq^2\,d\cos\theta} = a(q^2) + b(q^2) \cos\theta + c(q^2)
\cos^2\theta\,, \eea where the $q^2$-dependent functions are given by
\begin{align}
a(q^2) =& {G_F^2\tau_K\over 256\pi^3 m_K^3}  |V_{us}|^2 \lambda^{1/2}(m_K^2,q^2,m_\pi^2) \lambda^{1/2}(q^2,m_\ell^2,m_{\nu}^2)\nn\\
& \times \left[
\left( 1 - \frac{m_\ell^2+m_{\nu}^2}{q^2} \right) \vert h_0(q^2)\vert^2  +\left( \frac{m_\ell^2+m_{\nu}^2}{q^2} - \frac{(m_\ell^2-m_{\nu}^2)^2}{q^4} \right) \vert h_t(q^2)\vert ^2\right],\\
b(q^2) =& {G_F^2\tau_K\over 128\pi^3 m_K^3}  |V_{us}|^2 \lambda^{1/2}(m_K^2,q^2,m_\pi^2) \lambda(q^2,m_\ell^2,m_{\nu}^2)\frac{m_\ell^2-m_{\nu}^2}{q^4}\mathrm{Re}\left[ h_0(q^2)h_t^\ast(q^2)\right],\\
c(q^2) =& -{G_F^2\tau_K\over 256\pi^3 m_K^3}  |V_{us}|^2 \lambda^{1/2}(m_K^2,q^2,m_\pi^2) \frac{\lambda^{3/2}(q^2,m_\ell^2,m_{\nu}^2)}{q^4} \vert h_0(q^2)\vert^2\,.
\end{align}
After integrating over $\theta$ we obtain the usual expression for the differential branching fraction, which is shortly written as
\bea
\frac{d\cb (K\to \pi\ell\nu) }{dq^2 } = 2\,\left[ a(q^2) + \frac{1}{3}  c(q^2) \right]\,.
\eea
Finally, after integrating in $q^2$ and splitting up the pieces with contributions of massless and massive neutrinos in the final state, we have
\begin{align}\label{eq:BrKl3}
\cb (K\to \pi\ell\nu)= & (1- |U_{\ell 4}|^2) \int\displaylimits_{m_\ell^2}^{(m_K-m_\pi)^2}\left. \frac{d\cb (K\to \pi\ell\nu) }{dq^2 }\right|_{m_\nu=0} \nn\\
&+ \vartheta\left( m_K-m_\pi - m_\ell - m_4 \right) |U_{\ell 4}|^2 \int\displaylimits_{(m_\ell + m_4)^2}^{(m_K-m_\pi)^2} \frac{d\cb (K\to \pi\ell\nu_4) }{dq^2 }.
\end{align}
Another observable relevant to $K\to \pi \ell\nu$ decays can be easily
obtained after subtracting the number of events in the backward from the forward hemispheres. The
resulting forward-backward asymmetry is given by
\begin{align}\label{eq:Afb}
\!\!\! A_{\rm fb}^\ell (q^2) = {\displaystyle{\int_0^1  d\cos\theta \frac{d^2\cb (K\to \pi\ell\nu) }{dq^2\,d\cos\theta} } - \int_{-1}^0  d\cos\theta \displaystyle{\frac{d^2\cb (K\to \pi\ell\nu) }{dq^2\,d\cos\theta} }
\over \displaystyle{\int_{-1}^1 d\cos\theta \frac{d^2\cb (K\to \pi\ell\nu) }{dq^2\,d\cos\theta} } } \; .
\end{align}
Since there are three independent functions in the angular decay distribution~\eqref{eq:angular} we can define 
one more linearly independent observable, in addition to $d\cb/dq^2$ and $A_{\rm fb} (q^2)$. We choose the third observable to be the {charged} lepton polarization asymmetry.  
For that purpose we define the projectors $P_\pm = (1 \pm \slashed{s}\gamma_5)/2$ where the projection is made along the lepton polarization vector, 
\bea
s=\left(  \frac{ |\vec p_\ell |}{m_\ell},  \frac{E_\ell}{m_\ell}   \frac{\vec p_\ell}{ |\vec p_\ell | }\right)\,. 
\eea
The differential branching fraction can be separated into the positive lepton helicity and the negative one, i.e.
\begin{align}
\frac{d\cb (K\to \pi\ell\nu) }{dq^2 }& = \frac{d\cb_+ (K\to \pi\ell\nu) }{dq^2 } + \frac{d\cb_- (K\to \pi\ell\nu) }{dq^2 }, 
\end{align}
or, for short, $\cb = \cb_+ +\cb_-$. The lepton polarization asymmetry is then defined as
\begin{align}\label{eq:PL1}
P_\ell(q^2)  = { \quad \displaystyle{ \frac{d\cb_+  }{dq^2 } }-  \displaystyle{ \frac{d\cb_-   }{dq^2 }} \quad \over \displaystyle{ \frac{d\cb_+   }{dq^2 } } + \displaystyle{ \frac{d\cb_-   }{dq^2 } }} =& {1\over \displaystyle{(d\cb/dq^2)}} {G_F^2\tau_K\over 384 \pi^3 m_K^3}  |V_{us}|^2 \lambda^{1/2}(m_K^2,q^2,m_\pi^2) {\lambda(q^2,m_\ell^2,m_\nu^2)\over q^2} \nn\\
& \times \left[ \left( -2 + \frac{m_\ell^2-m_\nu^2}{q^2}\right) \vert h_0(q^2)\vert^2  + 3  \frac{m_\ell^2-m_\nu^2}{q^2}   \vert h_t(q^2)\vert^2  \right]\,,
\end{align}
or, in terms of form factors and by explicitly displaying the sum over the neutrino species, we write
\begin{align}\label{eq:PL}
P_\ell(q^2)  =& {1\over \displaystyle{(d\cb/dq^2)}} {G_F^2\tau_K\over 192 \pi^3 m_K^3}  |V_{us}|^2 {\lambda^{1/2}(m_K^2,q^2,m_\pi^2) \over q^4} \sum_{i=1}^4  |U_{\ell i}|^2 \lambda(q^2,m_\ell^2,m_{\nu_i}^2)  \nn\\
& \times \left[  \lambda(m_K^2,q^2,m_\pi^2) \left( \frac{m_\ell^2-m_{\nu_i}^2}{2 q^2} -1 \right) |f_+(q^2)|^2  +  \frac{3}{2} \frac{m_\ell^2-m_{\nu_i}^2}{ q^2} (m_K^2-m_\pi^2)^2 |f_0(q^2)|^2   \right].
\end{align}
Measuring $A_{\rm fb}^\ell (q^2)$ and $P_\ell(q^2)$ is hardly possible, but measuring the integrated characteristics might be feasible. This is why in the phenomenological application we will be using
$\langle A_{\rm fb}^\ell \rangle$ and $\langle P_\ell \rangle$, which are obtained by separately integrating the numerator and the denominator in both Eqs.~\eqref{eq:Afb} and~\eqref{eq:PL1}.

\subsection{Loop-induced weak decay $K\to \pi \nu \nu$}

Details of the derivation of the expressions for this decay rate can be found in Appendix~\eqref{app:a2} of the present paper. 
Here we only quote the corresponding effective Hamiltonian that we use, namely,
\bea\label{eq:Heff2}
\mathcal{H}_\eff = \frac{ \sqrt{2} G_F  \alpha_{\mathrm{em}}}{\pi}  \sum_{i,j=1}^4 {\widetilde C}_L^{ij}  \left(\bar s\gamma_\mu P_L d\right) \left( \nu_i  \gamma^\mu P_L \nu_j\right) +\mathrm{h.c.}
\eea
where 
\bea
\label{Ctilde}
\widetilde C_L^{i,j} = \frac{ 1 }{ \sin^2\theta_W} \sum_{\ell\in
  \{e,\mu,\tau\}} U_{\ell i}^\ast \left[ \lambda_c X_c^\ell +
  \lambda_t X_t \right] U_{\ell j}\,, \eea with $\lambda_c =
V_{cs}^\ast V_{cd}$, $\lambda_t = V_{ts}^\ast V_{td}$. The loop
contribution arising from the top quark amounts to
$X_t=1.47(2)$~\cite{Brod:2010hi}, while the box diagram with the
propagating charm depends on the lepton also in the loop, and yields $
X_c^e = X_c^\mu = 10.0(7)\times 10^{-4}$, $ X_c^\tau = 6.5(6)\times
10^{-4}$~\cite{Buras:1998raa}.  Notice also that the sum in the Wilson
coefficient $\widetilde C_L^{i,j}$ runs over the charged lepton species and
the one in Eq.~\eqref{eq:Heff2} over the neutrino mass
eigenstates. Using the same decomposition of the matrix element in
terms of the hadronic form factors, already defined in
Eq.~\eqref{eq:Kl3FF}, and assuming all neutrinos to be of Majorana
nature, we have
\begin{align}
\label{kpinunuMaj1}
\frac{d\cb (K^+\to \pi^+\nu\nu) }{dq^2 } = & \mathop{ \sum_{i,j=1}^{4}}_{i\leq j} \left( 1 - \frac{1}{2 }\delta_{ij} \right)  { \alpha_{\mathrm{em}}^2 G_F^2 \tau_{K^+} \over 768 \pi^5 m_K^3  }
 \lambda^{1/2}(m_K^2,q^2,m_\pi^2) {\lambda^{1/2}(q^2,m_{\nu_i}^2,m_{\nu_j}^2)\over q^2} \nn\\
 & \times \left\{  \vert \widetilde C_L^{ij}\vert^2 \left[ \lambda(m_K^2,q^2,m_\pi^2)\left( 2 - \frac{m_{\nu_i}^2 + m_{\nu_j}^2}{q^2}- \frac{(m_{\nu_i}^2 - m_{\nu_j}^2)^2}{q^4}\right) |f_+(q^2)|^2\right.
 \right.\nn\\
&\qquad \left. + 3 \left(  \frac{m_{\nu_i}^2 + m_{\nu_j}^2}{q^2}- \frac{(m_{\nu_i}^2 - m_{\nu_j}^2)^2}{q^4}\right) (m_K^2-m_\pi^2)^2 |f_0(q^2)|^2 \right] \nn\\
& \left. - 6\frac{m_i m_j}{q^2} \widehat C_L^{ij} \left[ \lambda(m_K^2,q^2,m_\pi^2) |f_+(q^2)|^2 - (m_K^2-m_\pi^2)^2 |f_0(q^2)|^2\right]\right\},
\end{align}
where 
\bea
\widehat C^{ij}_L =\frac{ 1 }{ \sin^4\theta_W}  \sum_{\ell,\ell^\prime\in \{e,\mu,\tau\}} \left(\lambda_c X_c^\ell +\lambda_t X_t \right) \left(\lambda_c X_c^{\ell'} +\lambda_t X_t \right)^\ast \times \mathrm{Re}\left[ U_{\ell i}^\ast U_{\ell j} U_{\ell' i}^\ast U_{\ell' j} \right] \,.
\eea
One should be particularly careful when using the above formula because the leptonic mixing matrix elements are in general complex, and while the functions $X_c^\ell$ and $X_t$ are real, the 
Cabibbo-Kobayashi-Maskawa couplings have both real and imaginary parts. More specifically, and by using the CKMfitter results~\cite{Charles:2004jd}, we obtain
\bea
\mathrm{Re}\lambda_t=-3.31(9)\times 10^{-4},\quad  
\mathrm{Re}\lambda_c=-0.2193(3), \quad
\mathrm{Im}\lambda_t=-\mathrm{Im}\lambda_c = 1.38(5)\times 10^{-4}.
\eea 
The above formula reduces to the Standard Model one after setting $m_{\nu_i}=m_{\nu_j}=0$, and by using the unitarity of the $4\times 4$ matrix.

\bigskip

If this decay occurs between the neutral mesons, the situation is slightly more delicate. When considering $K_L\to \pi^0 \nu\nu$, one should first keep in mind that $\vert K_L \rangle = ( \vert K^0\rangle +  \vert \overline K^0\rangle  )/\sqrt{2}$, 
which then means that the effective Hamiltonian~\eqref{eq:Heff2} between the initial and the final hadrons will result  
in two hadronic matrix elements which are related to each other by \textit{CP} symmetry, namely,
\bea
\langle \pi^0 | \bar d \gamma_\mu s|\overline K^0\rangle =  - \langle \pi^0 | \bar s \gamma_\mu d| K^0\rangle .
\eea
Furthermore, after invoking the isospin symmetry, we have
\bea
 \langle \pi^0 | \bar s \gamma_\mu d| K^0 \rangle =  \langle \pi^0 | \bar s \gamma_\mu u | K^+\rangle = \frac{1}{\sqrt{2}}  \langle \pi^+ | \bar s \gamma_\mu u | \overline K^0\rangle ,
\eea
where the last matrix element (to a charged pion) is the one defined in Eq.~\eqref{eq:Kl3FF}. With this, we can compute the decay rate and we obtain 
\begin{align}\label{Kpinunu}
\frac{d\cb (K_L\to \pi^0\nu\nu) }{dq^2 } = & \mathop{ \sum_{i,j=1}^{4}}_{i\leq j} \left( 1 - \frac{1}{2 }\delta_{ij} \right)  { \alpha_{\mathrm{em}}^2 G_F^2 \tau_{K_L} \over 768 \pi^5 m_K^3  }
 \lambda^{1/2}(m_K^2,q^2,m_\pi^2) {\lambda^{1/2}(q^2,m_{\nu_i}^2,m_{\nu_j}^2)\over q^2} \nn\\
 & \times \left\{  \vert \widetilde C_0^{ij}\vert^2 \left[ \lambda(m_K^2,q^2,m_\pi^2)\left( 2 - \frac{m_{\nu_i}^2 + m_{\nu_j}^2}{q^2}- \frac{(m_{\nu_i}^2 - m_{\nu_j}^2)^2}{q^4}\right) |f_+(q^2)|^2\right.
 \right.\nn\\
&\qquad \left. + 3 \left(  \frac{m_{\nu_i}^2 + m_{\nu_j}^2}{q^2}- \frac{(m_{\nu_i}^2 - m_{\nu_j}^2)^2}{q^4}\right) (m_K^2-m_\pi^2)^2 |f_0(q^2)|^2 \right] \nn\\
& \left. - 6\frac{m_i m_j}{q^2} \widehat C_0^{ij} \left[ \lambda(m_K^2,q^2,m_\pi^2) |f_+(q^2)|^2 - (m_K^2-m_\pi^2)^2 |f_0(q^2)|^2\right]\right\},
\end{align}
where 
\begin{align}
&\widetilde C_0^{i,j} =  -\frac{ 1 }{ \sin^2\theta_W} \sum_{\ell\in \{e,\mu,\tau\}} U_{\ell i}^\ast  \mathrm{Im}\left[ \lambda_c X_c^\ell + \lambda_t X_t \right] U_{\ell j}, \nn\\
&\widehat C^{ij}_0 =\frac{ 1 }{ \sin^4\theta_W}  \sum_{\ell,\ell^\prime\in \{e,\mu,\tau\}} \mathrm{Im}\left(\lambda_c X_c^\ell +\lambda_t X_t \right)   \mathrm{Im}\left(\lambda_c X_c^{\ell'} +\lambda_t X_t \right)   \mathrm{Re}\left[ U_{\ell i}^\ast U_{\ell j} U_{\ell' i}^\ast U_{\ell' j} \right] .
\end{align}
Like before, if we set $m_{\nu_i}=m_{\nu_j}=0$ and use the $U$ matrix unitarity, the above formula will lead to the familiar Standard Model expression (see e.g.~\cite{Buras:1998raa}). 

\subsection{``Invisible decay" $K_L\to \nu\nu$}

One might also look for an ``{\sl invisible decay}", such as the decay
of a kaon to neutrinos only.  We use the effective
Hamiltonian~\eqref{eq:Heff2}, and express the hadronic matrix element as
\bea \langle 0\vert \bar s\gamma_\mu \gamma_5 d\vert K_L(p)\rangle =
\langle 0\vert \bar d\gamma_\mu \gamma_5 s\vert K_L(p)\rangle=
\frac{i}{\sqrt{2}} f_Kp_\mu, \eea consistent with Eq.~\eqref{eq:fk},
and derive the expression for the decay rate by keeping in mind that
$CP|K^0\rangle = - |\overline K^0\rangle$. We obtain
\begin{align}\label{Knunu}
\cb (K_L\to \nu\nu)& = \mathop{ \sum_{i,j=1}^{4}}_{i\leq j}  \left( 1 - \frac{1}{2 }\delta_{ij} \right)  \frac{\alpha_{\rm em}^2 G_F^2 \tau_{K_L}}{8 \pi^3 m_{K}^3 \sin^4\theta_W} f_K^2 \lambda^{1/2}(m_{K}^2,m_{\nu_i}^2,m_{\nu_j}^2)\nn\\
& \times \left[  \biggl| \sum_{\ell\in \{e,\mu,\tau\}} \!\! \mathrm{Re}\left( \lambda_c X_c^\ell + \lambda_t X_t \right) U_{\ell i}^\ast U_{\ell j}\biggr|^2 \left( m_K^2 (m_{\nu_i}^2 + m_{\nu_j}^2 ) - (m_{\nu_i}^2 - m_{\nu_j}^2)^2 \right) \right. \nn\\
& \left.+\ 2\!\! \sum_{\ell,\ell'\in \{e,\mu,\tau\}} \!\! \mathrm{Re}( \lambda_c X_c^\ell + \lambda_t X_t ) \mathrm{Re}( \lambda_c X_c^{\ell'} + \lambda_t X_t  ) \mathrm{Re}( U_{\ell i}^\ast U_{\ell j} U_{\ell' i}^\ast U_{\ell' j} ) \; m_{\nu_i}  m_{\nu_j} m_K^2\right].
\end{align}
Since we consider the neutrinos to be Majorana fermions, the processes
$K\to \pi \nu\nu$ and $K_L\to \nu\nu$ can be viewed as lepton number
violating, and as such they can be used to probe the Majorana phases
via the last term in Eqs.~(\ref{Kpinunu}) and~(\ref{Knunu}). Notice,
however, that this term is multiplied by the product of neutrino
masses $m_{\nu_i}m_{\nu_j}$, and since in our scenario only one
neutrino can be massive the other ones are extremely light so that the
product of masses will be negligibly small.  The only nonzero
possibility is then $i=j=4$, but in this case the Majorana phases
cancel out in the product $U_{\ell 4}^\ast U_{\ell 4} U_{\ell' 4}^\ast
U_{\ell' 4}$. For this reason, the Majorana phases will not be
discussed in what follows. We should also note that Eq.~\eqref{Knunu},
with the appropriate simplifications, agrees with the one presented in
Ref.~\cite{Marciano:1996wy}.

\section{Results and discussion}\label{sec:discussion}

In this section we use the points selected by the constraints
discussed in Sec.~\ref{sec:scan} and evaluate the sensitivity of the
kaon decay observables on the presence of a massive sterile neutrino.
Whenever possible, and to make the situation clearer, for a given
observable $\mathcal{O}$ we will consider the ratio \bea
R_\mathcal{O}= \frac{\mathcal{O}}{\mathcal{O}^\mathrm{SM}}\,, \eea
where in the numerator we compute a given observable in the scenario
with three active and one massive sterile neutrino and divide it by its Standard Model prediction. Whenever possible,
those results will be compared with experimental values,
$R_\mathcal{O}^\mathrm{exp}$.
\vskip 4mm

\noindent
$\bullet$ We first examine the effects of sterile neutrinos on the leptonic decays of a charged kaon. To that end we define 
\bea
R_{K\ell 2} = \frac{\cb (K\to \ell \nu)}{\cb (K\to \ell \nu)^\mathrm{SM}}, 
\eea 
and compute its value by employing the expressions derived in the previous section.\footnote{The pioneering analysis of this ratio was made in~\cite{Shrock:1980vy,Shrock:1980ct}.} To estimate $R_{K\ell 2}^\mathrm{exp}$ we need an estimate of  $\cb (K\to \ell \nu)^\mathrm{SM}$, which we compute by 
using $|V_{us}|= 0.2255(4)$~\cite{Charles:2004jd},  $f_K=155.6(4)$~MeV computed in lattice QCD~\cite{Aoki:2016frl}, and by 
adding 
the electroweak and radiative corrections~\cite{Cirigliano:2011ny,Rosner:2015wva,Antonelli:2009ws}. We thus end up with
\begin{align}
\cb (K^+\to e^+ \nu)^\mathrm{SM} &= 1.572(10)\times 10^{-5},& R_{Ke 2}^\mathrm{exp} & =1.006(8)\,,\nn\\
\cb (K^+\to \mu^+ \nu)^\mathrm{SM} &= 63.55(39)\% ,& R_{K\mu 2}^\mathrm{exp}  &=0.999(6)\,,\nn\\
\cb (\tau^+\to K^+ \nu)^\mathrm{SM} &= 7.14(4)\times 10^{-3} ,& R_{\tau K 2}^\mathrm{exp}  &=0.980(15)\,,
\end{align}
where, in evaluating the $R_{K\ell 2}^\mathrm{exp}$ ratios, we used the {\it average} of the experimental results collected in Ref.~\cite{PDG}. 

\begin{figure}[ht!]
\centering
\hspace*{-6mm}\includegraphics[width=0.5\linewidth]{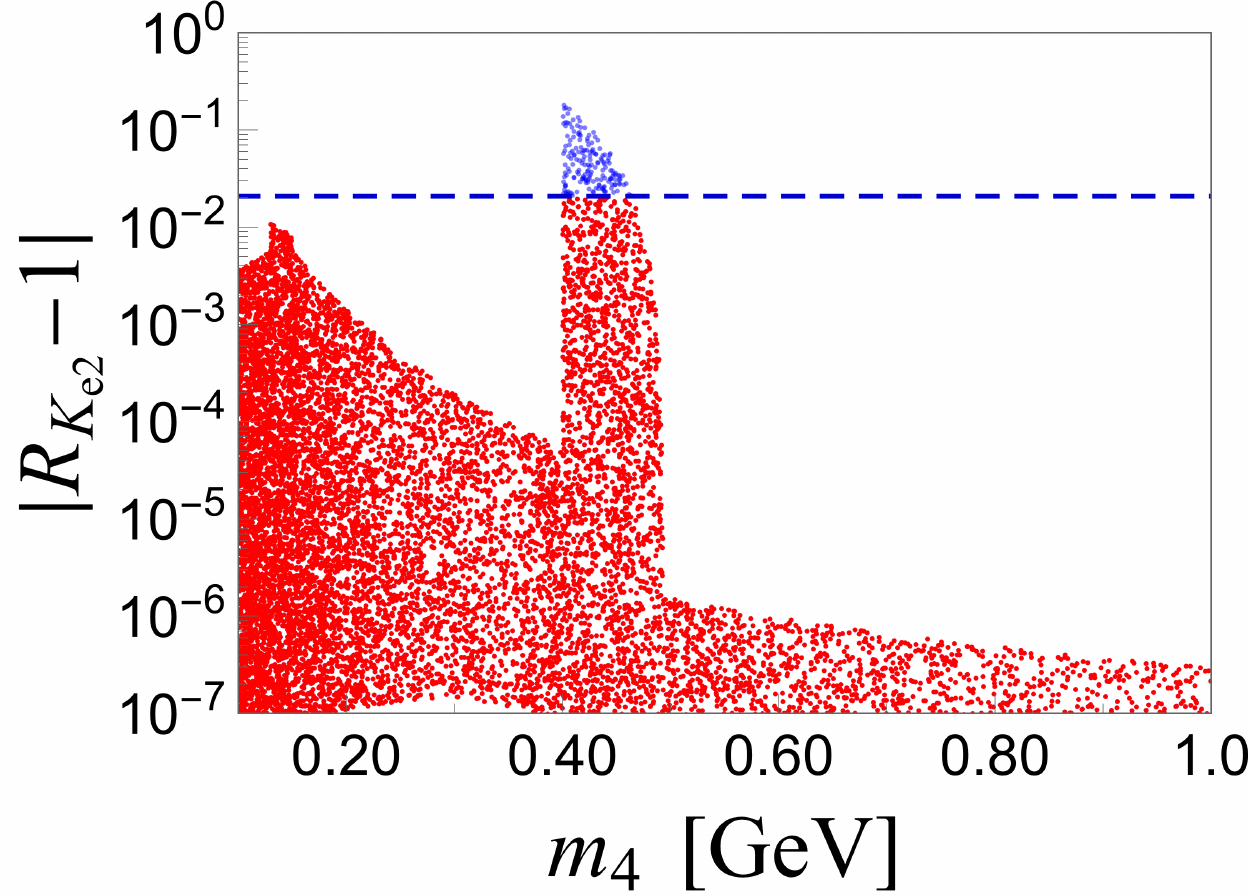}~\includegraphics[width=0.5\linewidth]{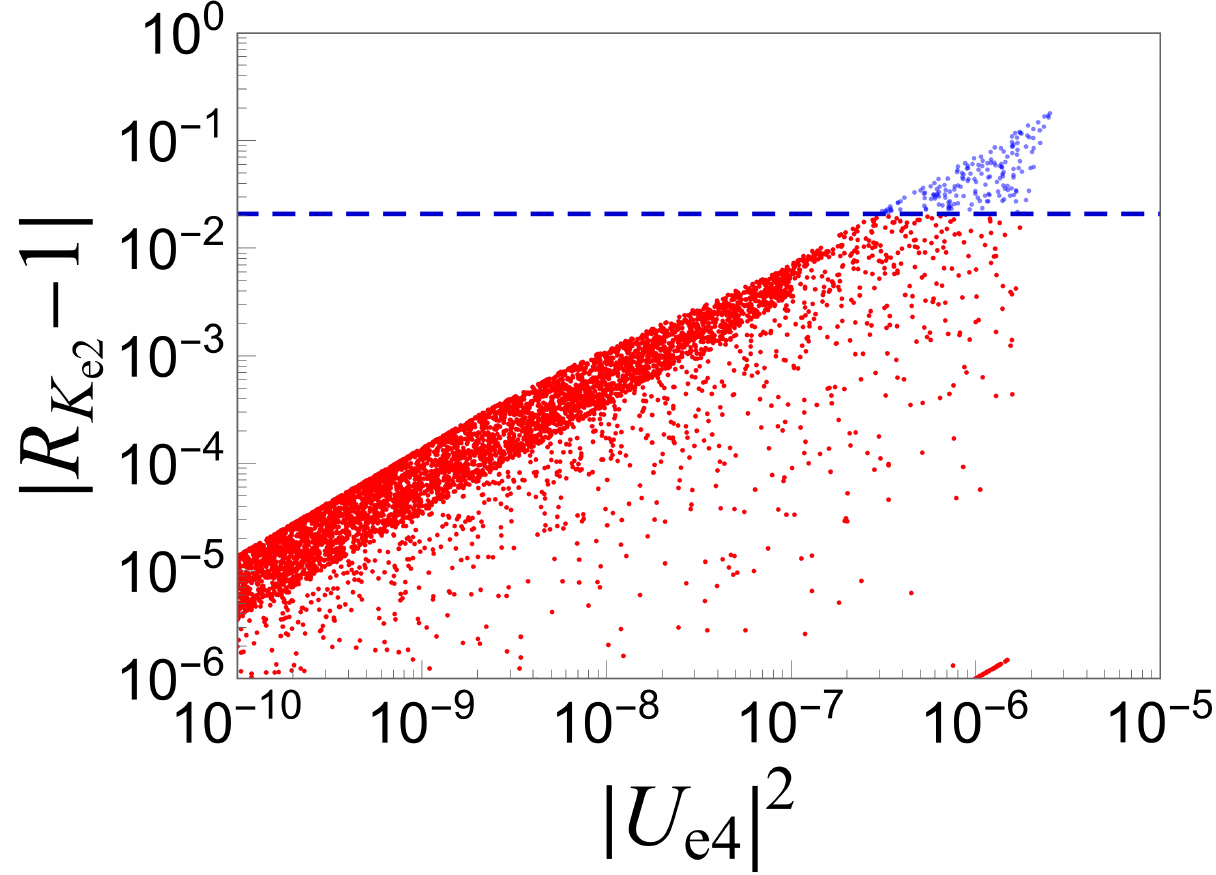}
\caption{\small \sl $|R_{Ke2}-1|$ as a function of the sterile
  neutrino mass, $m_4$ (left panel) and of $U_{e4}$ (right panel). The
  full ensemble of points (red and blue ones) correspond to $R_{Ke2}$
  computed using Eq.~\eqref{eq:BrKl2} in the Standard Model and in our
  scenario, with an additional sterile neutrino, by using the
  parameters selected in the scan discussed in
  Sec.~\ref{sec:scan}. Blue points are in conflict with $R_{Ke
    2}^\mathrm{exp}$ shown by the dashed line.}
\label{fig:3}
\end{figure}

Adding a massive sterile neutrino with parameters selected in a way
discussed in Sec.~\ref{sec:scan}, results in values of the branching
fractions which always fall within the experimental bounds except in
the case of the mode $K\to e \nu$, where some points get outside the
range allowed by experiment.  This situation is depicted in
Fig.~\ref{fig:3} where we see that requiring an agreement with the
experimental bound on $R_{Ke2}$ amounts to a new constraint in the region of $m_4\in (400\;\mev
,m_K)$.  In Fig.~\ref{fig:3} we also show the impact of
$R_{Ke2}^\mathrm{exp}$ on the corresponding active-sterile mixing
angle, or better $U_{e4}$.

This finding is actually equivalent to what has been discussed in Ref.~\cite{Abada:2013aba} where it has been shown that $\Delta r_K$ (defined analogously to Eq.~\eqref{rp} of the present paper) provides a useful constraint 
when building a viable extension of the Standard Model by including one (or more) sterile neutrino(s). Knowing that 
\bea
\Delta r_K = { R_{Ke 2}\over R_{K\mu 2} } - 1\,,
\eea
and since $R_{Ke 2}$ is currently constraining while $R_{K\mu 2}$ is not, it is clear that the two constraints are indeed equivalent. 
\vskip 4mm

\begin{figure}[t!]
\centering
\hspace*{-6mm}\includegraphics[width=0.5\linewidth]{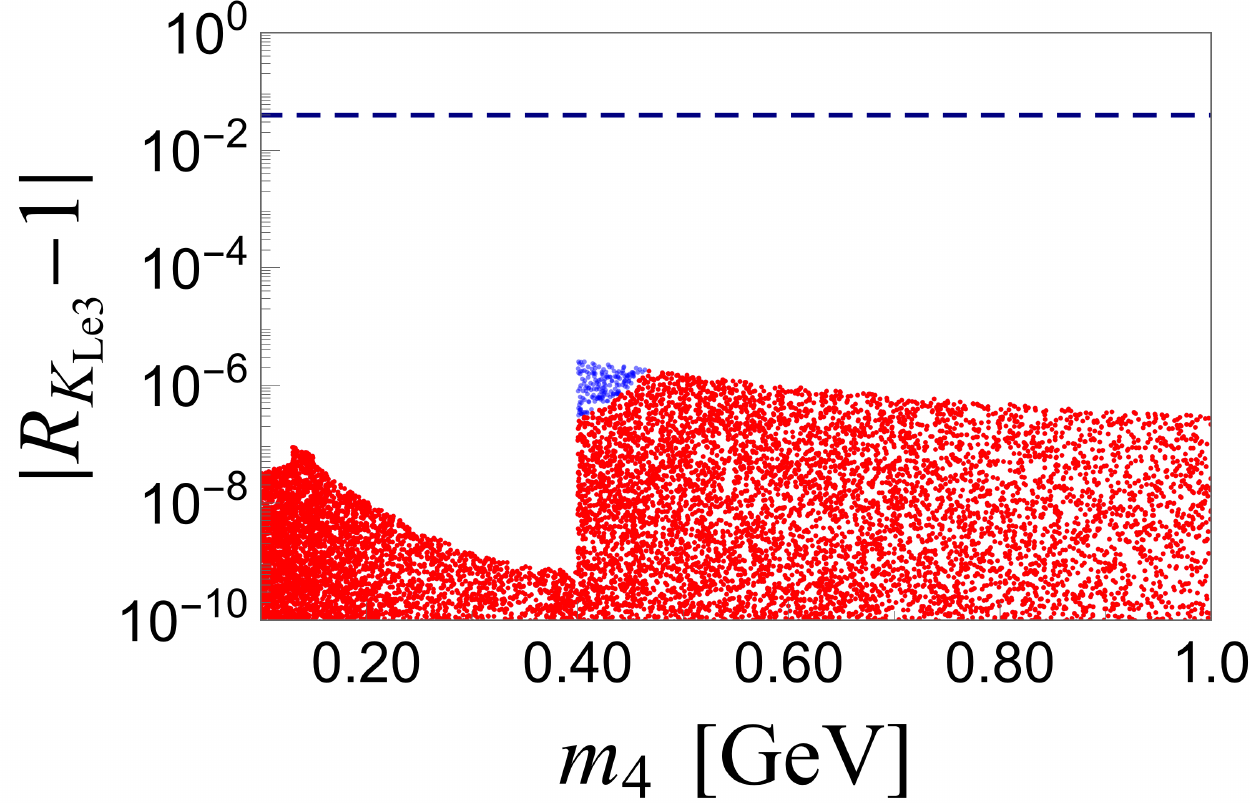}~\includegraphics[width=0.5\linewidth]{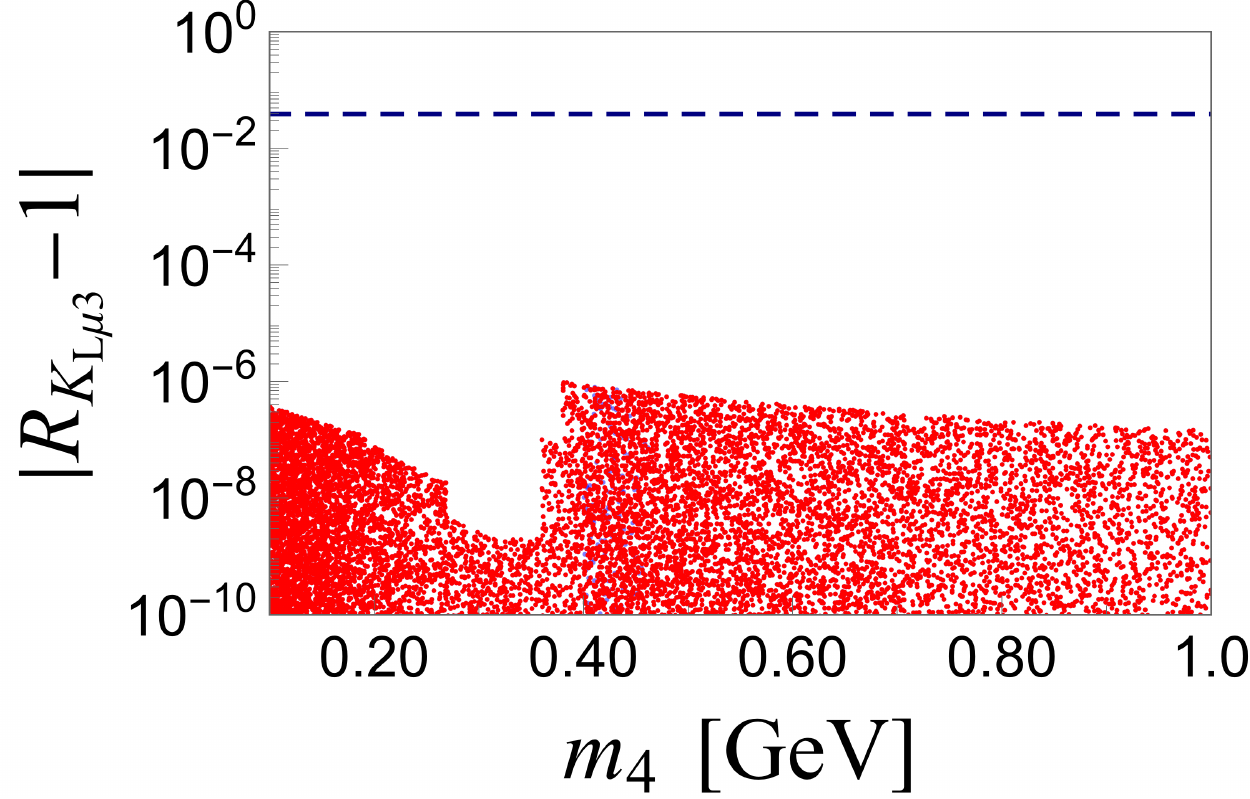}
\caption{\small \sl  Predictions for $|R_{Ke3}-1|$ and $|R_{K\mu 3}-1|$, computed by using the parameters obtained in Sec.~\ref{sec:scan}, are shown as functions of the mass of the sterile neutrino $m_4$. 
All the allowed points in red remain far below the experimental limit, shown by the dashed line. Blue points correspond to those discarded by incompatibility with $R_{Ke 2}^\mathrm{exp}$.}
\label{fig:4}
\end{figure}

\noindent
$\bullet$ As for the semileptonic decays, we focus on the decays of $K_L$ in order to avoid the uncertainties related to the isospin corrections which are present in the decays of charged kaons. 
The main remaining worry is to handle the hadronic uncertainties, i.e. those associated with the form factors $f_{+,0}(q^2)$. Those uncertainties are nowadays under control thanks to the recent 
precision lattice QCD computation with $N_\mathrm{f}=2+1+1$ dynamical quark flavors, presented in Ref.~\cite{Carrasco:2016kpy}. In that paper the authors computed the form factors at several $q^2$'s 
which are then fitted to the dispersive parametrization of Ref.~\cite{Bernard:2009zm}. We use those results in our computation and obtain,
\begin{align}
\cb (K_L\to \pi^- e^+ \nu)^\mathrm{SM} &=  41.31(46)\%,& R_{ Ke 3 }^\mathrm{exp} & =0.980(13)\,,\nn\\
\cb (K_L\to  \pi^- \mu^+ \nu)^\mathrm{SM} &= 27.51(29)\% ,& R_{ K\mu 3 }^\mathrm{exp}  &=0.981(11)\, ,
\end{align}
thus about $1.5\sigma$ away from the Standard Model prediction. Those bounds\footnote{{Although the mass of the sterile neutrino $m_4$ can in principle have
any value, we focused in Figs. ~\ref{fig:3} and ~\ref{fig:4} on the mass range $m_4\in [0,1]$~GeV. When the sterile neutrino is (not) kinematically accessible, we sum over all the (three) four neutrino final states;  beside, the effect of the presence of the sterile neutrino is also encoded in the modification of the neutral and charged current [see Eq.~(\ref{mod.current})], or equivalently in the effective coupling $g_A$. This effect is always present even if the sterile neutrino is not kinematically allowed, as one can see in Figs ~\ref{fig:3} and ~\ref{fig:4}.}}, however,
remain far too above the results we obtain after
including an extra sterile neutrino, which is also shown in
Fig.~\ref{fig:4}. In other words, the presence of a massive sterile
neutrino has a very little impact on the branching fractions of the
semileptonic kaon decays $K_{\ell 3}$. Even a significantly increased precision of
those measurements is very unlikely to unveil the presence of a sterile
neutrino in these decay modes.

As for the other two observables, we first 
computed them in the Standard Model and obtained
\begin{align}
&K_L\to \pi e \nu :& \langle A_\mathrm{fb}^e\rangle^\mathrm{SM} &= 8.5(1)\times 10^{-5}, &  \langle P_{e}\rangle^\mathrm{SM} &= 0.999(16) ,\nn\\
&K_L\to \pi \mu \nu :& \langle A_\mathrm{fb}^\mu\rangle^\mathrm{SM} &= 0.271(4), &  \langle P_{\mu}\rangle^\mathrm{SM} &= 0.088(5). 
\end{align}
We then checked their values in our scenario with one massive sterile neutrino and found that they change by a completely insignificant amount (at the one per-mil level). 
For example, we get 
\begin{align}
-2 \times 10^{-7} \leq& \left( \langle A_\mathrm{fb}^\mu\rangle - \langle A_\mathrm{fb}^\mu\rangle^\mathrm{SM}\right)\leq 5\times 10^{-7}\nn\\
-3.6 \times 10^{-6} \leq& \left( \langle P_\mu\rangle - \langle P_\mu\rangle^\mathrm{SM}\right)\leq 2.3\times 10^{-6}.
\end{align}
To understand why these quantities remain so insensitive to the presence of a heavy sterile neutrino, we checked all the constraints employed in our scan of parameters, and found that the most
severe constraints come from the direct searches, i.e. those we took from Ref.~\cite{Atre:2009rg}. Once taken into account, these constraints prevent the kaon physics observables from deviating from 
their Standard Model values. 
\vskip 4mm

\begin{figure}[t!]
\centering
\hspace*{-6mm}\includegraphics[width=0.5\linewidth]{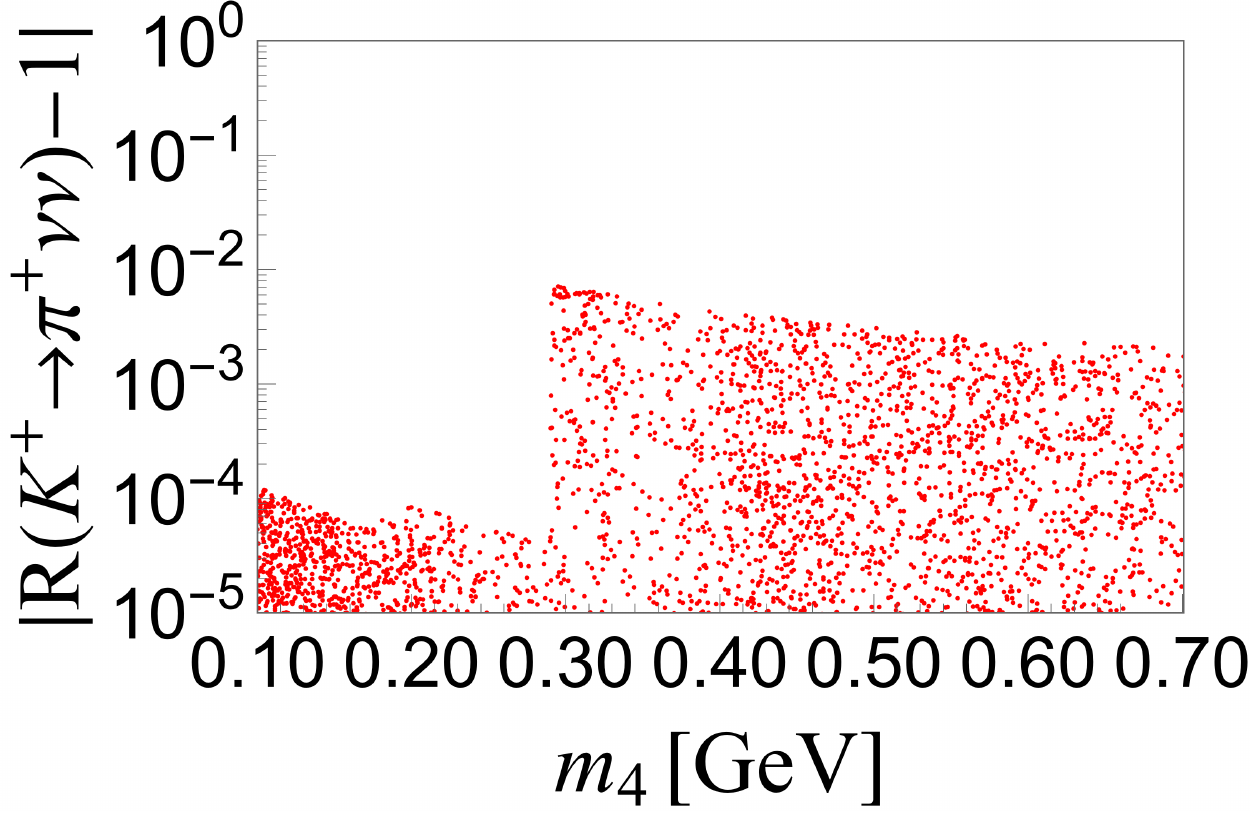}~\includegraphics[width=0.5\linewidth]{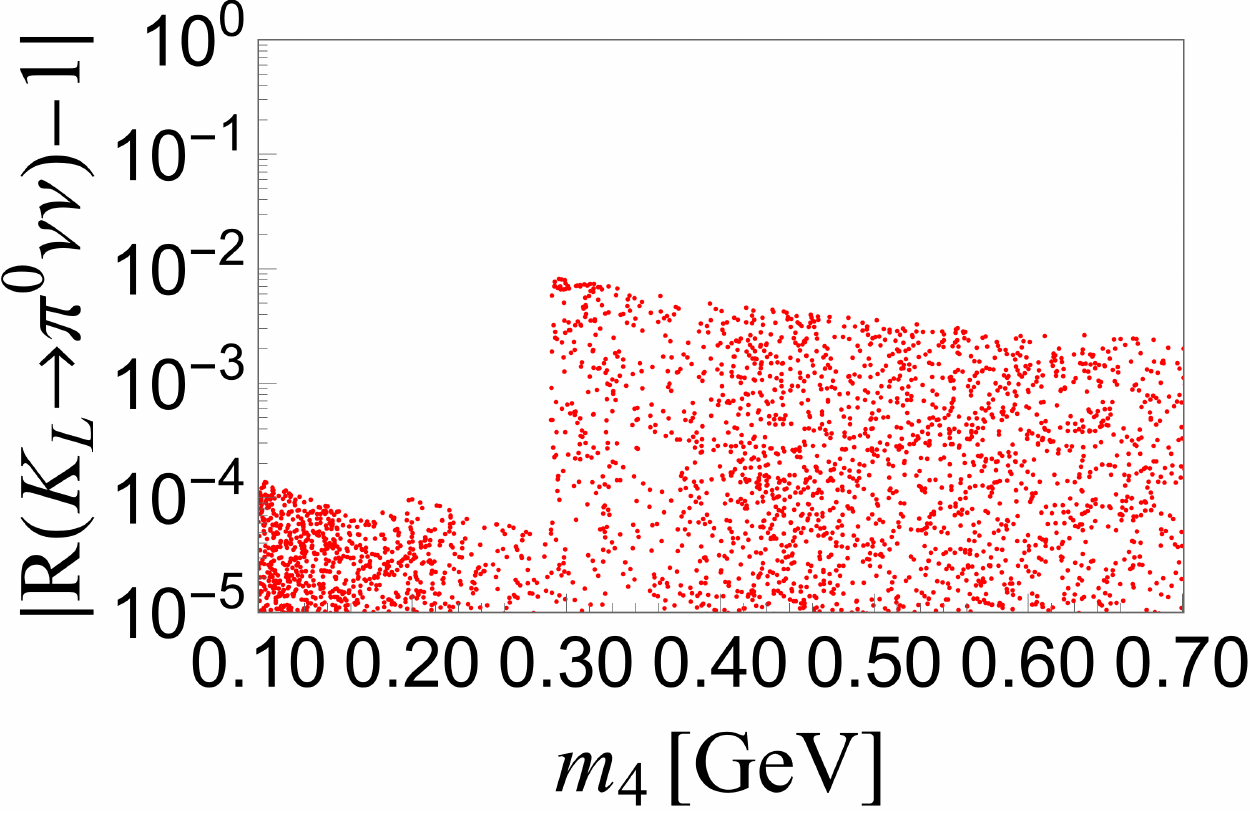}
\caption{\small \sl $|R_{K^\pm\to \pi^\pm \nu\nu}-1|$ and $|R_{K_L\to \pi^0 \nu\nu}-1|$ remain within $1\%$, which means that $K\to \pi \nu\nu$ decays are not sensitive to the presence of an extra (massive) sterile neutrino once experimental constraints are applied.}
\label{fig:5}
\end{figure}

\noindent
$\bullet$ The most interesting decay modes are expected to be the ones with two neutrinos in the final state. In the Standard Model, we have~\cite{Buras:2015qea}
\begin{align}\label{eq:SMnunu}
&\cb^{\mathrm{SM}}(K_L\to \pi^0\nu\nu) = (3.00\pm 0.30) \times 10^{-11},\nn\\ 
&\cb^{\mathrm{SM}}(K^\pm \to \pi^\pm\nu\nu) = (9.11\pm 0.72) \times 10^{-11},
\end{align}
where a control over the remaining long-distance hadronic contribution
to the charged mode can be achieved through numerical simulations of
QCD on the lattice for which a strategy has been recently developed in
Ref.~\cite{Christ:2016eae}.  These two decay modes are also subjects
of an intense experimental research at CERN (NA62) for the charged
mode~\cite{Moulson:2013grr}, and at J-PARC (KOTO) for the neutral
one~\cite{Ahn:2016kja}. We therefore find it important to examine in
which way their rates could be affected if the Standard Model is
extended by an extra sterile neutrino. It turns out that experimental constraints limit the deviation from 
the Standard Model prediction to less than $1\%$, which in view of the Standard Model uncertainties
[cf. Eq.~\eqref{eq:SMnunu}] means that the $K\to\pi \nu\nu$ decay
modes remain blind to the presence of an extra sterile neutrino. This
is illustrated in Fig.~\ref{fig:5}.
More specifically, we find 
\begin{align}
\forall \ m_4 \leq 1\ \gev\,,\qquad  R_{K_L\to \pi^0 \nu\nu} =1.000(9),\quad  R_{K^\pm \to \pi^\pm \nu\nu} =1.000(8)\,.
\end{align}
In other words, measuring $\cb(K_L\to \pi^0 \nu\nu)$ and $\cb(K^\pm \to \pi^\pm \nu\nu)$ consistent with the Standard Model predictions would be perfectly consistent with a scenario in which the Standard Model is extended by an extra sterile neutrino. 
Notice again that the cut into the parameter space in the region around $m_4 \approx 0.29$~GeV -- shown in Fig.~\ref{fig:5} -- comes from the direct searches~\cite{Atre:2009rg}, implemented in our scan.
\vskip 4mm 

\begin{figure}[t!]
\centering
\includegraphics[width=0.65\linewidth]{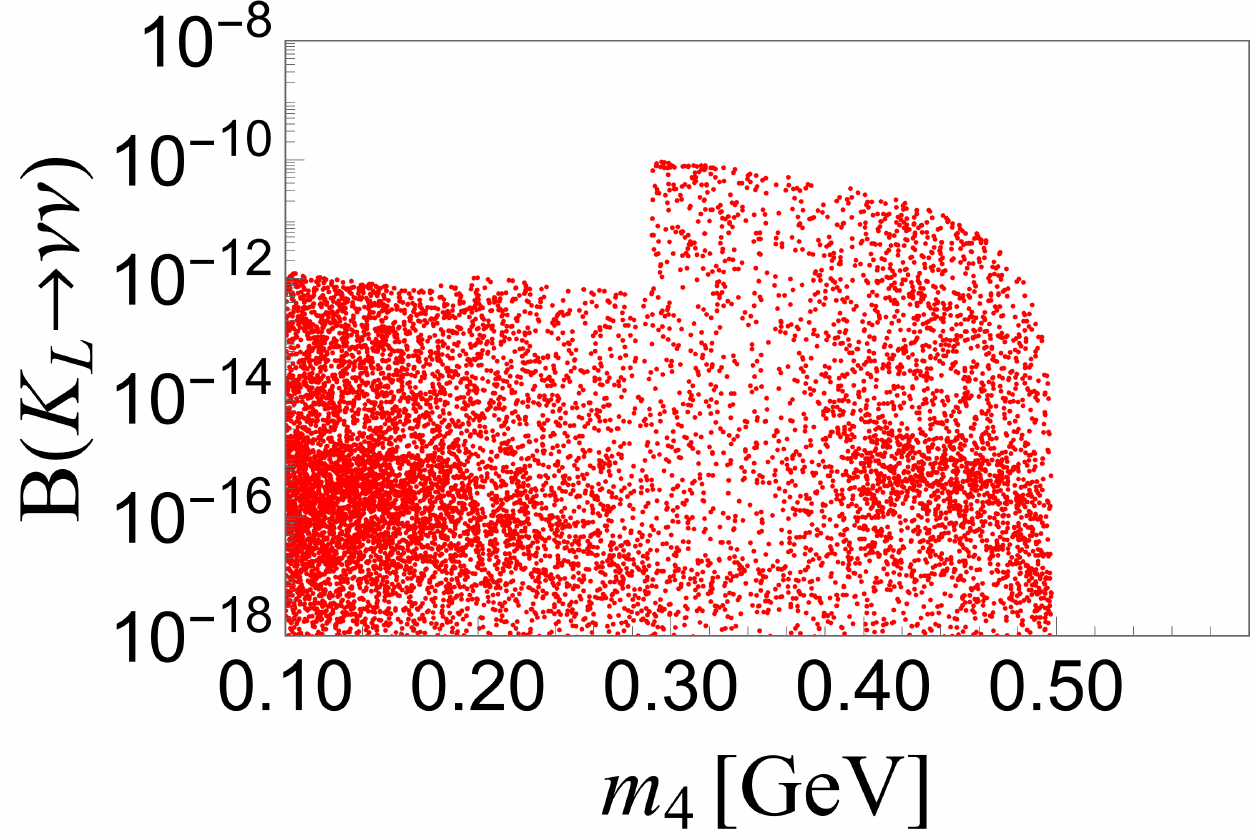}
\caption{\small \sl $ \cb(K_L\to \nu \nu) $ as a function of mass of the sterile neutrino for $m_4\leq m_{K^0}$. Notice that in the Standard Model
this branching fraction is zero while the values close to the upper bound 
found here are possibly within the reach of the KOTO, NA62(-KLEVER) and SHIP experiments. }
\label{fig:6}
\end{figure}

\noindent
$\bullet$ Finally, a similar analysis of the ``{\sl invisible kaon decay}"  $K_L\to \nu \nu$  shows that this mode can be largely enhanced if the sterile neutrino is massive. Due to the available phase space, this decay can be studied 
for $m_4\leq m_{K^0}$, and the result is shown in Fig.~\ref{fig:6}.
Knowing that in the Standard Model $\cb(K_L\to \nu \nu)^\mathrm{SM} \approx 
0$, the enhancement we observe is indeed substantial and since its
decay rate can be comparable to $\cb(K \to \pi\nu\nu)$ its
experimental research becomes highly important.
It has been proposed to search for this decay at the NA64 experiment
using $K_L$ produced from a $K^+$ beam hitting a target~\cite{NA64}.
From our analysis we
find the upper bound, 
 \bea \cb(K_L\to \nu \nu) \leq 1.2 \times
10^{-10}, \eea 
which could be within the reach of the KOTO, NA62(-KLEVER) and SHIP experiments even if the above bound is by an order
of magnitude lower.
The KOTO experiment aims to reach a sensitivity of $10^{-11}$ to $K_L \rightarrow \pi^0 \nu \nu$ in its first phase~\cite{Ahn:2016kja} and 
to have $4\times 10^{14}\,K_L$ at the entrance of the detector in phase 2~\cite{KOTOproposal}.
NA62-KLEVER is a project that would succeed NA62 and would aim to produce $3\times 10^{13}\,K_L$~\cite{Moulson:2016zsl}.
However, we would like to point out that the decay $K_L \to \nu\nu$ could also be searched for in $D$-meson decays, making use of the relatively large branching ratio
$\cb(D \to K_L \pi)\sim 1\%$ and tagging the $K_L$ via the pion. Beam dumps experiments at the CERN SPS like SHIP or a possible run of NA62 in a beam dump configuration
would produce copious amounts of $D$ mesons. A year of running in beam dump mode for NA62 would produce $\sim 10^{15}\,D$ mesons~\cite{NA62dump}, which would correspond to roughly
$10^{13}$ tagged $K_L$. SHIP would accumulate even more data, producing $6.8\times 10^{17}\,D$ mesons~\cite{Alekhin:2015byh}, which would translate into more than $7\times 10^{15}$ tagged $K_L$.
In any case, an experimental bound on this decay
mode would be of great importance for studying the effects of physics
beyond the Standard Model in the leptonic sector. Obviously, a
nonzero measurement of $\cb(K_L\to \nu \nu)$ would be a clean signal
of the non-Standard Model physics.

Before closing this section, we should make a brief comment on the
lepton flavor violating kaon decays, which in our scenario would be generated by the heavy neutrino running in the loop. By
using the formulas given in Ref.~\cite{Becirevic:2016zri} 
trivially adapted to the kaon decays, and the result of the scan of
Sec.~\ref{sec:scan}, we obtain that these modes are completely
negligible, i.e. the branching fractions of all these modes are under
$10^{-16}$.~\footnote{To be more specific, we get \bea \cb (K_L\to\mu
  e)< 10^{-18},\quad \cb (\tau\to K_S\mu )< 10^{-16},\quad \cb
  (\tau\to \phi \mu )< 10^{-16},\quad \cb (K^+\to \pi^+\mu e)<
  10^{-23}.\nn \eea}

\section{Summary }
\label{sec:conc}

In this paper we presented the results of our study concerning the
impact of a massive sterile neutrino on the weak kaon decays such as
the leptonic, semileptonic and the decay of a kaon to neutrinos.  In
the effective approach adopted in this work, one sterile neutrino is
supposed to mimic the effect of a more realistic model in which the
neutrino sector is extended to include one or more sterile neutrinos.

Although the mass of the sterile neutrino $m_4$ can in principle have
any value, we focused on the mass range $m_4\in [0,1]$~GeV, and in particular 
on $m_4\leq m_K$, when the sterile neutrino is kinematically accessible. 
In order to constrain six new parameters ($m_4$, the
three sterile-active neutrino mixing angles and two new phases) we
used a number of quantities discussed in the body of the paper,
together with the perturbative unitarity requirement, as well as the
constraints arising from the direct searches~\cite{Atre:2009rg}.
After combining such selected parameters with the expressions for the
leptonic and semileptonic decays we derive here, we found that only
$\mathcal{B}(K\to e\nu )$ can significantly deviate from the current experimental
value.  That conflict with the data is present in the interval $m_K >
m_4 \gtrsim 0.4$~GeV.  The other quantities, including the forward
backward and the lepton polarization asymmetries, remain unchanged
with respect to their Standard Model values with the effect of the massive
sterile neutrino remaining at the level of less than $1\%$.

We also derived the expressions for the kaon decays to two (Majorana)
neutrinos in the final state, namely $\cb(K \to \pi \nu\nu)$ and
$\cb(K_L \to \nu\nu)$.  Our expressions are generic and can be used
when studying a new physics scenario in which heavy neutrinos with no new gauge couplings are
involved. This will be increasingly relevant with the ongoing
experimental effort at CERN (NA62, NA64, SHIP) and J-PARC (KOTO) targeting
$\mathcal{B}(K^\pm \to \pi^\pm \nu\nu)$ and $\mathcal{B}(K_L \to \pi^0
\nu\nu)$, respectively.  These two decays, however, appear to be
insensitive to the massive sterile neutrino once the experimental and theoretical constraints are taken into account.  In other words, if the
experimental results of the weak kaon decays turn out to be consistent
with the Standard Model predictions to a $1\%$ uncertainty, this would
not be in contradiction with the neutrino sector extended by a
massive and relatively light sterile neutrino(s).  The only kaon decay mode which appears
to be sensitive to the presence of a massive sterile neutrino is $K_L \to \nu\nu$,
the branching fraction of which can go up to $\mathcal{O}( 10^{-10})$,
thus possibly within reach of the NA62(-KLEVER), SHIP and KOTO experiments. Knowing that the Standard
Model value of this mode is zero, its observation would be a clean
signal of new physics.

Notice also that $\cb(K \to \pi \nu\nu)$ and $\cb(K_L \to \nu\nu)$
could be used to probe the Majorana phases in the models in which more
than one massive neutrino is considered.  In the approach adopted in
this paper only one neutrino can be heavy and therefore such a study
is prohibited. If instead one considers a realistic model with more
than one heavy neutrino then a study of the Majorana phases becomes
possible too~\cite{inPREPA}.

We should mention that one can also consider the situation with a very
heavy sterile neutrino ${\mathcal O}(1\,\mathrm{TeV})$. In that case
the processes discussed in this paper could be modified by the effects
of violation of the mixing matrix unitarity; see for instance\cite{Blennow:2016jkn}. We checked that
possibility in the explicit computation and found that such effects
are indeed tiny.  Importantly, however, a heavy sterile neutrino can
propagate in the loop-induced processes and shift the values of
$\cb(K_S\to \mu\mu)$ and $\cb(K\to \pi \mu\mu)$.  We checked that the
corresponding effect remains small and completely drowned in the large
(long-distance QCD) uncertainties already present in the Standard
Model estimates of $\cb(K_S\to \mu\mu)$ and $\cb(K\to \pi
\mu\mu)$~\cite{Antonelli:2009ws,Mescia:2006jd}.

Finally, the expressions presented in this paper can be easily
extended to other similar decays, such as $D$-, $D_s$-, $B$- and
$B_s$-meson decays. We decided to focus on the kaon decays because
of the recent theoretical developments in taming the hadronic
uncertainties and because of the better experimental precision.
\vskip 1.5cm
\noindent 
{\bf  Acknowledgments:} {\small \sl 
We thank  S. Gninenko (spokesperson of NA64) for very interesting  remarks.This project has received funding from the European Union's Horizon 2020 research and innovation program under the Marie Sklodowska-Curie Grants No. 690575 and No. 674896. C.W. receives financial support from the European Research
Council under the European Union's Seventh Framework Programme
(Grants No. FP/2007-2013)/ERC Grant NuMass Grant No. 617143.  C.W.~thanks the Universit\'e Paris Sud, the University of T\"ubingen and the IBS Center for Theoretical Physics of the Universe for their 
hospitality during the final stages of this project.
R.Z.F. thanks the Universit\'e Paris Sud for the kind hospitality, and CNPq and FAPESP for partial financial support. }

\begin{appendix}

\section{Majorana vs Dirac case}

Being electrically neutral neutrinos can be described by using either Dirac or Majorana spinors. A Majorana spinor obeys the condition
\begin{equation}
\label{MajoranaCondition}
 \psi=\psi^C = \xi\, \mathcal{C} \bar \psi^T\,,
\end{equation}
where $C$ denotes the charge conjugation and $\xi$ an arbitrary phase factor that can be absorbed into redefinition of $\psi$. 
As a consequence, a Majorana spinor has only 2 degrees of freedom (whereas a Dirac spinor has 4) and can be expressed
using only the left-handed (LH) chiral component
\begin{equation}
 \psi= \psi_L + (\psi_L)^C\,.
\end{equation}
In the massless limit the chiral components decouple and Dirac and
Majorana spinors verify the same Dirac equation. Since in the Standard
Model only the LH component is subject to gauge interactions, Dirac
and Majorana neutrinos will behave identically in the massless limit
and only the observables that exhibit a dependence on the neutrino
mass can probe the nature of
neutrinos. This behavior is at the core of the \textit{confusion
  theorem}~\cite{Kayser:1981nw}.

The plane wave expansion of a Dirac spinor reads
\begin{equation}
 \psi_\mathrm{Dirac} (x)= \int \frac{d^3 p}{\sqrt{(2\pi)^3 2 E_p}} \sum_s \left[ a_s(p) u_s(p) e^{-\imath p.x} + b_s^\dagger (p) v_s(p) e^{\imath p.x} \right] \,,
\end{equation}
and that of a Majorana spinor has the following form, 
\begin{equation}
\label{Majorana wave}
 \psi_\mathrm{Majorana} (x)= \int \frac{d^3 p}{\sqrt{(2\pi)^3 2 E_p}} \sum_s \left[ a_s(p) u_s(p) e^{-\imath p.x} + \xi a_s^\dagger (p) v_s(p) e^{\imath p.x} \right] \,,
\end{equation}
where $a_s^\dagger(p)$ and $b_s^\dagger (p)$ are the particle and antiparticle creation operators, while $u_s(p)$ and $v_s(p)$ are the positive and negative energy spinors. 
Since a theory with Majorana fermions does not conserve fermion number, multiple spinor contractions are allowed, leading to possible ambiguities
in calculations.

\subsection{Feynman Rules\label{app:a1}}

A possible modification of the Feynman rules to account for Majorana fermions was presented in~\cite{Denner:1992vza}, nowadays widely used and also incorporated in automated 
tools like \texttt{FeynArts}~\cite{Hahn:2000kx} or \texttt{Sherpa}~\cite{Gleisberg:2008ta,Hoche:2014kca}, for example. 
Before giving the list of vertices that we used and their expressions for Majorana neutrinos, let us illustrate the difference between vertices for Dirac and Majorana neutrinos.

In the weak basis, the $Z\nu\nu$ interactions are flavor diagonal and given for both Majorana and Dirac neutrinos by
\begin{equation}
 \mathcal{L}_{Z\nu\nu}= -\frac{g_2}{2 \cos{\theta_W}} \bar \nu'_{Li} \slashed Z P_L \nu'_{Li}\,.
\end{equation}
The neutrino mass matrix $M$ is put in a diagonal form using a singular value decomposition for Dirac neutrinos
\begin{equation}
 U_L^\dagger M U_R = \mathrm{diag}(m_{\nu_i})\,,
\end{equation}
where $i=1,...,n$ with $n$ the number of neutrinos and
\begin{equation}
 \nu'_L=U_L P_L \nu\,, \quad \nu'_R= U_R P_R \nu\,,
\end{equation}
and the lepton mixing matrix $U$ is given by
\begin{equation}
U_{\alpha i}\, =\,\sum_{k=1}^{3} V^*_{k\alpha}\, U_{L_{ki}}\,.
\end{equation}
In the case of Majorana neutrinos, the neutrino mass matrix $M$ is diagonalized using Eq.~(\ref{RotDiag}).
For both Majorana and Dirac neutrinos, the $Z\nu\nu$ interactions are given by, in the mass basis,
\begin{equation}
 \mathcal{L}_{Z\nu\nu}= -\frac{g_2}{2 \cos{\theta_W}} \bar \nu \slashed Z U^\dagger U P_L \nu\,.
\end{equation}
Using the usual Feynman rules for Dirac neutrinos gives the following. \\
\vspace*{0.2cm}\\
\begin{minipage}{0.4\textwidth}
\begin{center}
\begin{fmffile}{Znunu1}
\fmfcmd{%
 style_def charged_boson expr p = 
 draw (wiggly p);
 fill (arrow p)
 enddef;}
\fmfset{arrow_len}{0.4cm}\fmfset{arrow_ang}{15}
\begin{fmfgraph*}(40,25)
\fmflabel{$Z$}{Z}
\fmflabel{$\nu_j$}{ni}
\fmflabel{$\nu_i$}{nj}
\fmfleft{Z}
\fmfright{ni,nj}
\fmfforce{(0.5w,0.5h)}{v}
\fmf{wiggly}{Z,v}
\fmf{fermion}{ni,v}
\fmf{fermion}{v,nj}
\end{fmfgraph*}
\end{fmffile}
\end{center}
\end{minipage} \hfill
\begin{minipage}{0.55\textwidth}
\begin{equation}
 - \frac{\imath g_2}{2 \cos \theta_W} \gamma_\mu (U^\dagger U)_{ij} P_L
\end{equation}
\end{minipage}\\
\vspace*{0.2cm}\\

However, we would expect the Feynman rule to exhibit some symmetry under the exchange $i \leftrightarrow j$ for Majorana neutrinos. Using the following properties of Majorana bispinors
\begin{eqnarray}
 \bar \psi \gamma_\mu \chi = - \bar \chi \gamma_\mu \psi\,,\qquad \bar \psi \gamma_\mu \gamma_5 \chi = \bar \chi \gamma_\mu \gamma_5 \psi\,,
\end{eqnarray}
we can rewrite the $Z\nu\nu$ interaction term for Majorana neutrinos in the mass basis as
\begin{equation}
  \mathcal{L}_{Z\nu\nu}=-\frac{g_2}{4 \cos \theta_W} \bar \nu \slashed Z \left[ (U^\dagger U) P_L - (U^\dagger U)^* P_R \right] \nu\,,
\end{equation}
which, keeping in mind that two contractions of the Majorana spinors are possible, gives the Feynman rule\\
\vspace*{0.2cm}\\
\begin{minipage}{0.4\textwidth}
\begin{center}
\begin{fmffile}{Znunu2}
\fmfcmd{%
 style_def charged_boson expr p = 
 draw (wiggly p);
 fill (arrow p)
 enddef;}
\fmfset{arrow_len}{0.4cm}\fmfset{arrow_ang}{15}
\begin{fmfgraph*}(40,25)
\fmflabel{$Z$}{Z}
\fmflabel{$\nu_j$}{ni}
\fmflabel{$\nu_i$}{nj}
\fmfleft{Z}
\fmfright{ni,nj}
\fmfforce{(0.5w,0.5h)}{v}
\fmf{wiggly}{Z,v}
\fmf{plain}{ni,v}
\fmf{plain}{v,nj}
\end{fmfgraph*}
\end{fmffile}
\end{center}
\end{minipage} \hfill
\begin{minipage}{0.55\textwidth}
\begin{equation}
 - \frac{\imath g_2}{2 \cos \theta_W} \gamma_\mu \left[ (U^\dagger U)_{ij} P_L - (U^\dagger U)_{ij}^* P_R \right]\,,
\end{equation}
\end{minipage}
\\
\vspace*{0.2cm}\\
\noindent which agrees with the prescription of~\cite{Denner:1992vza} and can be obtained as well by writing the matrix element
\begin{equation}
 \langle \nu_j \nu_i | \left( \frac{-g_2}{2 \cos \theta_W} \right) \bar \nu_\alpha \slashed Z (U^\dagger U)_{\alpha\beta} P_L \nu_\beta |0 \rangle\,,
\end{equation}
and doing the Wick contractions in all possible ways.  

On the opposite, due to the presence of a charged lepton that imposes a distinction between leptons and antileptons, the vertices involving a $W^\pm$ gauge boson are identical between 
Majorana and Dirac neutrinos,\\
\vspace*{0.2cm}\\
\begin{minipage}{0.4\textwidth}
\begin{center}
\begin{fmffile}{Wlnu1}
\fmfcmd{%
 style_def charged_boson expr p = 
 draw (wiggly p);
 fill (arrow p)
 enddef;}
\fmfset{arrow_len}{0.4cm}\fmfset{arrow_ang}{15}
\begin{fmfgraph*}(40,25)
\fmflabel{$W^-$}{W}
\fmflabel{$l_\alpha$}{li}
\fmflabel{$\nu_i$}{nj}
\fmfleft{W}
\fmfright{li,nj}
\fmfforce{(0.5w,0.5h)}{v}
\fmf{wiggly}{W,v}
\fmf{fermion}{v,li}
\fmf{plain}{v,nj}
\end{fmfgraph*}
\end{fmffile}
\end{center}
\end{minipage} \hfill
\begin{minipage}{0.55\textwidth}
\begin{equation}
 - \frac{\imath g_2}{\sqrt{2}} \gamma_\mu U_{\alpha i} P_L
\end{equation}
\end{minipage}
\\
\vspace*{1.3cm}\\
\begin{minipage}{0.4\textwidth}
\begin{center}
\begin{fmffile}{Wlnu2}
\fmfcmd{%
 style_def charged_boson expr p = 
 draw (wiggly p);
 fill (arrow p)
 enddef;}
\fmfset{arrow_len}{0.4cm}\fmfset{arrow_ang}{15}
\begin{fmfgraph*}(40,25)
\fmflabel{$W^+$}{W}
\fmflabel{$l_\alpha$}{li}
\fmflabel{$\nu_i$}{nj}
\fmfleft{W}
\fmfright{li,nj}
\fmfforce{(0.5w,0.5h)}{v}
\fmf{wiggly}{W,v}
\fmf{fermion}{li,v}
\fmf{plain}{v,nj}
\end{fmfgraph*}
\end{fmffile}
\end{center}
\end{minipage} \hfill
\begin{minipage}{0.55\textwidth}
\begin{equation}
 - \frac{\imath g_2}{\sqrt{2}} \gamma_\mu U^*_{\alpha i} P_L
\end{equation}
\end{minipage}
\\

\subsection{Detailed expression for $K^+\to \pi^+ \nu \nu$\label{app:a2}}

In this section, we present the complete analytical expressions for $K^+\to \pi^+ \nu \nu$ calculated for both Majorana and Dirac neutrinos and show that they are equivalent in the limit of massless
neutrinos, as expected from the confusion theorem. The expression for the kaon decays are found in Eq.~(\ref{kpinunuMaj1}) for Majorana neutrinos. In the case of Dirac neutrinos, we obtain
\begin{align}
\label{kpinunuDir1}
\left.\frac{d\cb (K^+\to \pi^+\nu\nu) }{dq^2 }\right|_{\mathrm{Dirac}} \!\!\! = & \mathop{ \sum_{i,j=1}^{4}} { \alpha_{\mathrm{em}}^2 G_F^2 \tau_{K^+} \over 1536 \pi^5 m_K^3  }
 \lambda^{1/2}(m_K^2,q^2,m_\pi^2) {\lambda^{1/2}(q^2,m_{\nu_i}^2,m_{\nu_j}^2)\over q^2} \nn\\
 & \times \vert \widetilde C_L^{ij}\vert^2 \left[ \lambda(m_K^2,q^2,m_\pi^2)\left( 2 - \frac{m_{\nu_i}^2 + m_{\nu_j}^2}{q^2}- \frac{(m_{\nu_i}^2 - m_{\nu_j}^2)^2}{q^4}\right) |f_+(q^2)|^2\right.
 \nn\\
&\qquad \left. + 3 \left(  \frac{m_{\nu_i}^2 + m_{\nu_j}^2}{q^2}- \frac{(m_{\nu_i}^2 - m_{\nu_j}^2)^2}{q^4}\right) (m_K^2-m_\pi^2)^2 |f_0(q^2)|^2 \right],
\end{align}
where $\widetilde C_L^{ij}$ is given in Eq.~(\ref{Ctilde}), while $f_{+,0}(q^2)$ are the form factors defined in Eq.~\eqref{eq:Kl3FF}.

Using the fact that for Majorana neutrinos,
\begin{equation}
 \frac{d\cb (K^+\to \pi^+\nu_i\nu_j) }{dq^2 }=\frac{d\cb (K^+\to \pi^+\nu_j\nu_i) }{dq^2 }\,,
\end{equation}
we can rewrite the branching ratio summed over all neutrinos in Eq.~(\ref{kpinunuMaj1}) as
\begin{align}
 \frac{d\cb (K^+\to \pi^+\nu\nu) }{dq^2 }   &= \mathop{ \sum_{i,j=1}^{4}}_{i\leq j} \left( 1 - \frac{1}{2 }\delta_{ij} \right) \frac{d\cb (K^+\to \pi^+\nu_i\nu_j) }{dq^2 } \nonumber \\
					    &= \frac{1}{2 } \mathop{ \sum_{i,j=1}^{4}}_{i<j} \left(\frac{d\cb (K^+\to \pi^+\nu_i\nu_j) }{dq^2 } + \frac{d\cb (K^+\to \pi^+\nu_j\nu_i) }{dq^2 } \right) 
						+ \frac{1}{2 } \mathop{ \sum_{i=1}^{4}} \frac{d\cb (K^+\to \pi^+\nu_i\nu_i) }{dq^2 } \nonumber \\
					    &= \frac{1}{2 } \mathop{ \sum_{i,j=1}^{4}} \frac{d\cb (K^+\to \pi^+\nu_i\nu_j) }{dq^2 }\,,
\end{align}
where we used the exchange symmetry $\nu_i\leftrightarrow \nu_j$ in the last equality. Therefore, 
\begin{align}
\label{kpinunuMaj2}
\left. \frac{d\cb (K^+\to \pi^+\nu\nu) }{dq^2 }\right|_{\mathrm{Majorana}} & \!\!\!  =  \mathop{ \sum_{i,j=1}^{4}}  { \alpha_{\mathrm{em}}^2 G_F^2 \tau_{K^+} \over 1536 \pi^5 m_K^3  }
 \lambda^{1/2}(m_K^2,q^2,m_\pi^2) {\lambda^{1/2}(q^2,m_{\nu_i}^2,m_{\nu_j}^2)\over q^2} \nn\\
 & \times \left\{  \vert \widetilde C_L^{ij}\vert^2 \left[ \lambda(m_K^2,q^2,m_\pi^2)\left( 2 - \frac{m_{\nu_i}^2 + m_{\nu_j}^2}{q^2}- \frac{(m_{\nu_i}^2 - m_{\nu_j}^2)^2}{q^4}\right) |f_+(q^2)|^2\right.
 \right.\nn\\
&\qquad \left. + 3 \left(  \frac{m_{\nu_i}^2 + m_{\nu_j}^2}{q^2}- \frac{(m_{\nu_i}^2 - m_{\nu_j}^2)^2}{q^4}\right) (m_K^2-m_\pi^2)^2 |f_0(q^2)|^2 \right] \nn\\
& \left. - 6\frac{m_i m_j}{q^2} \widehat C_L^{ij} \left[ \lambda(m_K^2,q^2,m_\pi^2) |f_+(q^2)|^2 - (m_K^2-m_\pi^2)^2 |f_0(q^2)|^2\right]\right\}\,.
\end{align}
It is then clear that the difference between Eqs.~(\ref{kpinunuDir1}) and~(\ref{kpinunuMaj2}) comes from the last term in Eq.~(\ref{kpinunuMaj2}), which is proportional to $m_i m_j$. As a 
consequence, the branching ratios for Majorana neutrinos and for Dirac neutrinos are equal in the limit of massless neutrinos, providing a concrete example of the confusion theorem.

\end{appendix}

\end{document}